\documentclass[jcp,aps,showpacs,supergroupedaddress,epsfig,amsmath,amssymb,twocolumn]{revtex4}
\usepackage{epsfig,amsmath,amsthm,amssymb,bm,epsf,graphics,psfrag,verbatim,subfigure,framed}
\usepackage[makeroom]{cancel}
\usepackage{mathtools}
\usepackage[usenames,dvipsnames]{color}
\usepackage{empheq}
\usepackage{bbm}
\usepackage{mathrsfs}
\usepackage{amsfonts}
\usepackage{color} 
\usepackage{pbox}
\usepackage{xcolor} 
\usepackage{multirow}
\def\nv{\mathbf{n}}
\def\mv{\mathbf{m}}
\def\Nv{\mathbf{N}}
\def\Mv{\mathbf{M}}

\newcommand{\be}{\begin{equation}}
\newcommand{\ee}{\end{equation}}
\newcommand{\ba}{\begin{eqnarray}}
\newcommand{\ea}{\end{eqnarray}}
\newcommand{\bw}{\begin{widetext}}
\newcommand{\ew}{\end{widetext}}

\newcommand{\rv}{{\mathbf{r}}}

\begin{document}

\title{Atom-surface interaction induced by quenched monopolar charge disorder}
\author{Bing-Sui Lu}
\email{binglu@aus.edu}
\affiliation{College of Arts and Sciences, American University of Sharjah, P.O. Box 26666, Sharjah, United Arab Emirates}
\date{\today}

\begin{abstract}
We study the modification to the energy level shifts of an atom induced by the quenched monopolar charge disorder inside the bulk of neighboring dielectric slabs as well as their surfaces. By assuming that the charge disorder follows Gaussian statistics with a zero mean, we find that the disorder generally results in a downward shift of the energy levels, which corresponds to an attractive force that can compete with and overcome the nonresonant Casimir-Polder force for sufficiently large atom-surface separations $z_0$. For an atom near a single semi-infinite slab with bulk (surface) charge disorder, the shift decays as $z_0^{-1}$ ($z_0^{-2}$). 
For both surface and bulk disorder, the shift is proportional to the variance of the charge disorder density.  
In addition, we investigate the behavior of the charge disorder-induced energy level shift for an atom confined to a vacuum gap between two coplanar and semi-infinite slabs of the same dielectric material, finding that the position of net zero disorder-induced force occurs closer to the surface of the slab with the smaller charge disorder variance. 
\end{abstract}

\maketitle

\section{Introduction}

Precision studies of the Casimir-Polder interaction~\cite{casimir1948,wylie1985,bloch2005,laliotis2021} and experiments probing hypothetical forces~\cite{mostepanenko2004,robertson2006,behunin2014a} require the elimination of ``parasitic" interactions that can mask the forces being measured, particularly if those forces are especially weak. A common source of such parasitic interactions arises from electrostatic disorder. In the context of Casimir~\cite{garrett2015} and Casimir-Polder force experiments~\cite{bloch2005,laliotis2021}, electrostatic disorder can take the form of a spatially random surface potential of the exposed surface of the material under measurement, which occurs if the material is polycrystalline with different grains having different surface potentials~\cite{ashcroft1976,dubessy2009}. Electrostatic disorder can also take the form of impurity charges that are randomly distributed in the bulk and/or the surface of the material~\cite{dean2012,pitaevskii2008,dalvit2008}. Such impurity charges can appear as a result of the fabrication process, or due to the adsorption of contaminants.
Electrostatic disorder can lead to deviations in Casimir-Polder measurements from theoretical predictions~\cite{chan2025}, a broadening of atomic energy levels~\cite{carter2011}, and limit the accuracy of detecting violations of Newton's inverse square law~\cite{mostepanenko2004,robertson2006,behunin2014a}. 
Besides being relevant to the precise measurement of weak forces, the study of charge disorder can be relevant to the transport of atoms along atom waveguides and hollow fibers~\cite{renn1995,dowling1996,vorrath2010}, as the mutual interplay of the charge disorder-induced force and the Casimir-Polder force~\cite{gorza2001,gorza2006,lu2022,lu2025a,lu2025b} can affect the mean trajectory of  atoms between the confining surfaces. 

\

With regard to the interaction between material surfaces, there have been studies on the effect of a random surface potential~\cite{camp1991,speake1996,behunin2014a,ke2023,speake2003,behunin2014b,behunin2012,fosco2013}, as well as studies on the effect of random bulk and surface monopolar charges~\cite{naji2005,podgornik2006,naji2010,sarabadani2010,dean2011,rezvani2012}.  
We are thus motivated to investigate the force on an atom induced by the presence of random quenched monopolar charges in the bulk as well as the surface of a planar dielectric substrate. 
A related study is Ref.~\cite{carter2011}, which pertains to the broadening of the energy levels of a Rydberg atom induced by surface patches with random potentials. The authors calculated the variance of the {\em first}-order perturbation correction to the atomic energy levels induced by the random patch potentials. 
On the other hand, as we are interested in the correction to the Casimir-Polder shift induced by the charge disorder, we have calculated this correction using {\em second}-order perturbation theory, as the first-order perturbation correction vanishes under averaging over the charge disorder statistics. As we focus on small atoms, we neglect multipolar terms higher than the dipolar one in the Hamiltonian. 

\

In this paper, we investigate the atom-surface interaction arising from the type of monopolar charge disorder which is {\em quenched}~\cite{ziman1979}, which means that the disorder does not relax on experimental timescales. Quenched disorder is quite prevalent and not limited to electrostatic ones, but occur in many other physical domains as well, such as DNA sequences~\cite{lu2015,lu2016} and liquid crystalline polymer networks~\cite{lu2011,lu2012,lu2013}. 
Our paper is organized as follows. In Sec.~II, we introduce a model of quenched charge disorder, and obtain an expression which relates the energy level shift in an atom induced by monopolar charge disorder to the atom's polarizability and the disorder-induced electric field. In Sec.~III, we study the behavior of the energy shift induced by the surface and bulk charge disorders for the following two scenarios: (i)~an atom near a semi-infinite dielectric slab, and (ii)~an atom confined to the vacuum gap between two coplanar and semi-infinite slabs, which we assume to be made of the same dielectric material. 
For the first scenario, we also compare the behavior of the disorder-induced force with that of the nonresonant Casimir-Polder force, which can be of relevance to Casimir-Polder force measurements. The second scenario can be relevant to atoms confined in rectangular hollow cores and atom waveguides~\cite{nha1996}. 
To connect with the pre-existing literature and for simplifying the appearance of the equations, we have expressed our results using cgs units.
Finally, in Sec.~IV, we illustrate our results using the example of a metastable helium atom in the $n=2$ triplet state near vitreous ${\rm SiO_2}$, which is an insulator. 

 
\section{The model}

\subsection{Quenched charge disorder} 

\begin{figure}[h]
    \centering
      \includegraphics[width=0.43\textwidth]{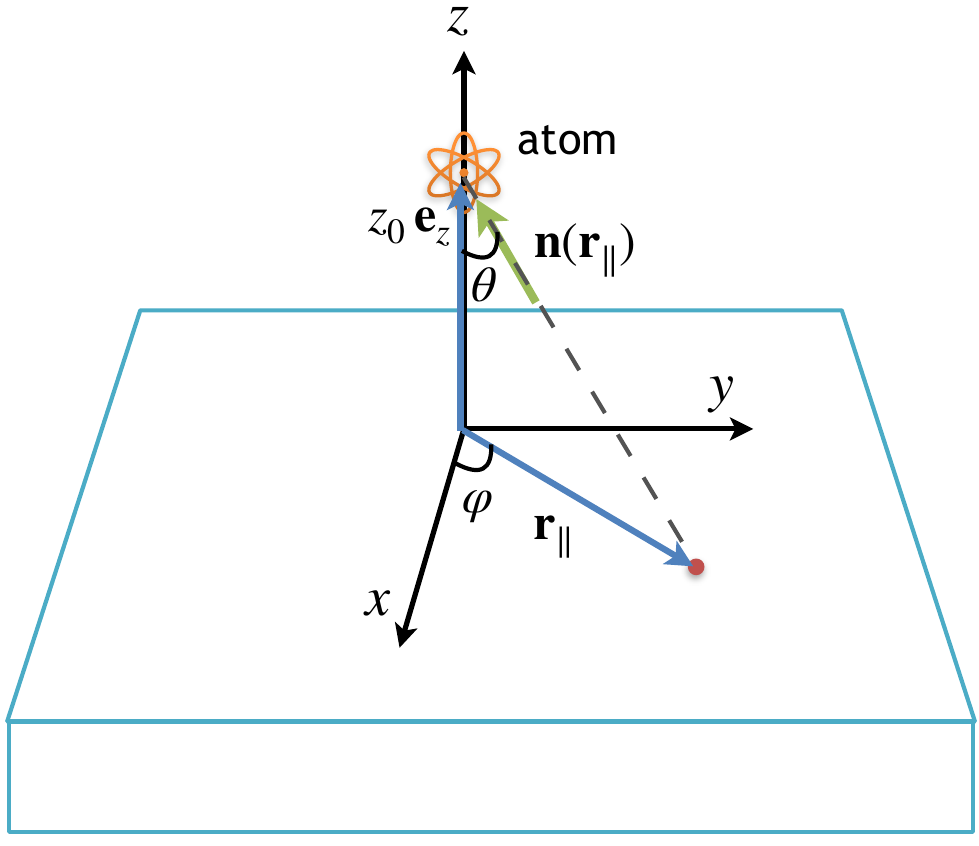}
      \caption{The atom next to a single semi-infinite slab: the atom is at $\rv = z_0 \, {\bf e}_z$, with the origin of the coordinate system being the point on the surface directly underneath the atom. The red dot at position $\rv_{\parallel}$ represents a given local surface charge disorder. The green arrow represents the direction of the unit vector ${\bf n}(\rv_{\parallel})$, which points from the local charge disorder at $\rv_{\parallel}$ to the atom. Not shown is the rest of the charge disorder, which can be in the bulk and/or the surface.} 
      \label{fig:geometry}
\end{figure} 
For an atom positioned at a height $z_0$ above the top surface of a semi-infinite planar dielectric slab~(see Fig.~\ref{fig:geometry}), we set the position of the top surface of the slab at $z=0$. Furthermore, we assume the quenched monopolar charge disorder is homogeneously distributed, which can occur inside the bulk of the slab as well as the surface. 
We also assume that each impurity charge has an equal probability of being positive or negative, so that the mean charge disorder is zero, thus ensuring the slab is overall charge neutral. The statistics of the charge disorder is described by 
\ba
\overline{\delta\rho(\rv)} &=& 0, 
\\
\overline{\delta\rho(\rv) \delta\rho(\rv')} &=& \mathcal{C}(\rv-\rv'), 
\label{C}
\ea
where the overline $\overline{\ldots}$ denotes an average over all realizations of the quenched disorder, and $\mathcal{C}(\rv-\rv')$ is a correlation function which is maximum when $\rv = \rv'$ and decays to zero as $|\rv-\rv'|\to\infty$. 

\

In what follows, we assume that the charge disorder is spatially uncorrelated, so we can express the correlation function in terms of a Dirac delta-function. 
This assumption corresponds to looking at atom-surface separations much greater than the correlation length of the monopolar charge disorder. In order to estimate the range of separations over which such an assumption is valid, we assume that the material is polycrystalline, with the surface potentials randomly varying across the grains. As similar impurity charges tend to be adsorbed onto the same grain, the grain size sets the correlation length of the charge disorder. If we consider an average grain size of 10 nm, the assumption of spatial decorrelation will be valid for separations much greater than 10 nm, which is the range that we consider in the present paper. 

\

Writing $\delta\rho_S$ and $\delta\rho_B$ for the surface and bulk charge disorder respectively, the charge disorder statistics is 
\ba
\overline{\delta\rho_S(\rv)} 
&\!\!=\!\!& 
0, \,\, 
\overline{\delta\rho_B(\rv)}
= 0, 
\nonumber\\
\overline{\delta\rho_S(\rv) \delta\rho_S(\rv')} 
&\!\!=\!\!& 
\sigma_S^2 \, \delta(\rv_{\parallel}-\rv_{\parallel}') \delta(z-0-)
\delta(z'-0-), 
\nonumber\\
\overline{\delta\rho_B(\rv) \delta\rho_B(\rv')} 
&\!\!=\!\!&  \sigma_B^2 \, \delta(\rv - \rv'). 
\label{var-oneslab}
\ea
Here, $\sigma_S^2$ denotes the mean-square charge disorder per unit area (or the variance of the surface charge disorder density), and $\sigma_B^2$ denotes the variance of the bulk charge disorder per unit volume. The symbol ``$0-$" indicates that the surface charge disorder resides just below $z=0$. 

\subsection{Charge disorder-induced energy shift}

The atom is polarized by the local field which is a superposition of two types of fluctuating fields: a field generated by the fluctuating dipoles inside the dielectric slab, as well as a field generated by the quenched and randomly distributed impurity charges. 
As we mentioned earlier, we assume that the atom is small, so that we can neglect the quadrupolar interaction, and describe the atom-slab interaction for a {\it given} realization of the charge disorder by the dipole interaction Hamiltonian:
\be
H_I = \hat{\mu}_{i} (\mathcal{E}_{0i}(\rv_0) +\delta \hat{\mathcal{E}}_{i}(\rv_0)), 
\label{HI}
\ee
where $i = x,y,z$ labels the Cartesian axis directions, and $\rv_0$ is the position vector of the atom in three-dimensional space. We take the origin to be the point on the slab surface directly beneath the atom, so that 
\be
\rv_0 = z_0 \, {\bf e}_z,
\ee
where ${\bf e}_z$ is the unit vector directed in the positive z-direction. 
The symbol $\delta \hat{\bm{\mathcal{E}}}(\rv_0)$ denotes the (quantum) electromagnetic field fluctuation at the atom's position, 
and ${\bm{\mathcal{E}}}_0(\rv_0)$ denotes the electric field generated at the atom's position by the monopolar impurity charges. 
For the case where the atom is near a single slab, and the impurity charges are on the slab's surface, the disorder-induced field at $\rv_0$ is given by 
\be
{\bm{\mathcal{E}}}_0(\rv_0) = \frac{2}{\varepsilon(0)+1} \int d^3\rv \, \frac{\delta \rho_S(\rv) \, \nv(\rv)}{|\rv - \rv_0|^2},  
\label{Ev0c}
\ee
where $\nv(\rv)$ is the unit vector directed from a given local charge disorder at position $\rv$ to the atom, and the surface charge disorder density is defined by 
\be
\delta\rho_S(\rv) \equiv \sum_{I} q_{I} \delta(\rv_{\parallel} - \rv_{I,{\parallel}})\delta(z), 
\ee
$\varepsilon(0)$ is the static dielectric permittivity of the slab, $q_{I}$ is the $I$th monopolar impurity charge, $\rv_{\parallel}$ is a two-dimensional position vector in the plane of the surface, and $\rv_{I,{\parallel}}$ is the two-dimensional position vector of the $I$th impurity charge in the plane of the surface. 
The vanishing of the mean charge disorder thus implies the vanishing of the disorder-averaged electric field, {\it viz}., 
\be
\overline{{\bm{\mathcal{E}}}_0(\rv_0)} = 0, 
\label{Ev0av}
\ee 
and a similar result holds for the case where the field is induced by impurity charges in the bulk. 

\

The dipole interaction in Eq.~(\ref{HI}) leads to a shift in the energy levels of the atom. As the dipole interaction is weak, we can treat the interaction Hamiltonian in Eq.~(\ref{HI}) as a perturbation. The first-order perturbation correction to the energy of the $n$th level is given by 
\ba
\delta E_n^{(1)} &=& \sum_N P(N) \langle N| \langle n| \hat{\mu}_{i} (\mathcal{E}_{0i}+\delta \hat{\mathcal{E}}_{i}) | n \rangle | N \rangle 
\\
&=& \mathcal{E}_{0i} \langle n|\hat{\mu}_{i} | n\rangle 
+  \sum_N P(N)  \langle N| \delta \hat{\mathcal{E}}_{i} |N\rangle \langle n|\hat{\mu}_{i} | n\rangle. 
\nonumber
\ea
In the equation above, $\{ |n\rangle \}$ is the set of eigenstates of the unperturbed atom, $\{ |N\rangle \}$ is the set of eigenstates of the unperturbed radiation bath, and $P(N)$ is the Boltzmann probability of finding the radiation bath to be in the eigenstate $|N\rangle$. 
The first term on the right-hand side of the above equation vanishes under disorder averaging (cf. Eq.~(\ref{Ev0av})), and the second term vanishes under thermal and quantum averaging. Thus, we consider the second-order perturbation correction, which is given by 
\ba
&&\delta E_n^{(2)} 
\!\!=\!\!
- \mathcal{P} 
\Bigg( 
\sum_{m,M,N} P(N) 
\frac{\langle N | \langle n | \hat{\mu}_{i} (\mathcal{E}_{0i}+\delta \hat{\mathcal{E}}_{i}) | m \rangle | M \rangle 
}{E_M - E_N + E_{m} - E_{n}} 
\nonumber\\
&&\qquad\times 
\langle M | \langle m | \hat{\mu}_{j} (\mathcal{E}_{0j}+\delta \hat{\mathcal{E}}_{j}) | n \rangle | N \rangle 
\Bigg).
\ea
After averaging over the charge disorder, we obtain
\ba
\label{full-shift}
&&\overline{\delta E_n^{(2)}}
\\
&\!\!=\!\!&
-\frac{1}{\hbar} 
\sum_{m} 
\frac{\overline{\mathcal{E}_{0i} \mathcal{E}_{0j}} \, \mu_{i}^{nm} \mu_{j}^{mn}}{\omega_{mn}} 
\nonumber\\
&&-\frac{1}{\hbar} \mathcal{P} 
\Bigg(
\sum_{m,M,N} P(N) 
\frac{\langle N | \delta \hat{\mathcal{E}}_{i} | M \rangle \langle M | \delta \hat{\mathcal{E}}_{j} | N \rangle \mu_{i}^{nm} \mu_{j}^{mn}}{\omega_M - \omega_N + \omega_{mn}} 
\Bigg). 
\nonumber
\ea
Here, $\omega_M \equiv E_M/\hbar$, $\omega_{mn} \equiv (E_m - E_n)/\hbar$ (where 
$\hbar = 1.0546\times 10^{-27} \, {\rm erg.s}$), $\mu_{i}^{nm} \equiv \langle n | \hat{\mu}_{i} | m \rangle$, and $\mathcal{P}$ means that we take the principal value. 
In the expression above, there are no cross-terms involving products of $\mathcal{E}_{0i}$ and $\delta \hat{\mathcal{E}}_{j}$, as  $\mathcal{E}_{0i}$ vanishes under disorder averaging. 
The first term on the right-hand side (RHS) represents the additional energy level shift induced by the presence of the charge disorder, whilst the second term leads to the Casimir-Polder shift for a charge disorder-free surface~\cite{wylie1985,gorza2001,gorza2006,laliotis2021,lu2022}. 
Using the formula for the polarizability of the $n$th atomic state~\cite{wylie1985}, 
\be
\alpha_{ij}^n(\omega) = \frac{2}{\hbar} \sum_m \frac{\omega_{mn} \mu_{i}^{mn} \mu_{j}^{nm}}{\omega_{mn}^2 - (\omega + i \eta)^2}, 
\ee
we can write the first term on the RHS of Eq.~(\ref{full-shift}) as 
\be
\delta E_n^{{\rm Stark}} = - \frac{1}{2} \alpha_{ij}^n (0) \overline{\mathcal{E}_{0i} \mathcal{E}_{0j}}.  
\label{energy-correction}
\ee
This correction describes a quadratic Stark effect produced by the charge disorder-induced electric field. Although we have derived the expression for the case of surface charge disorder, it applies equally well for the case of bulk charge disorder. 
The force generated by the charge disorder is then 
\be
f = - \frac{\partial \delta E_n^{{\rm Stark}}}{\partial z_0}. 
\ee

\subsection{Nonresonant Casimir-Polder shift} 

As we will subsequently compare the charge disorder-induced interaction correction with the nonresonant Casimir-Polder energy shift, let us recall the formula for the nonresonant Casimir-Polder shift of a metastable or ground-state atom near a surface at zero temperature, which includes retardation effects~\cite{wylie1985}: 
\be
\delta E^{CP}(\rv_0) = -\frac{\hbar}{2\pi} \int_0^\infty \!\!\! d\xi \, G_{ij}^{R} (\rv_0, \rv_0; i\xi) \alpha_{ij}(i\xi).    
\ee
We consider a metastable helium atom in the $n=2$ triplet state (also known as ${{\rm He^\ast}}$)~\cite{bruhl2002,bordag2009}, whose dynamic polarizability can be described by the following one-oscillator model,   
\be
\alpha_{ij}(i\xi) = \frac{\alpha_0}{1 + (\xi/\omega_0)^2} \delta_{ij},
\label{alpha-He*}
\ee
with the static polarizability $\alpha_0 = 315.63 \, {\rm a.u.} = 4.678\times 10^{-23} \, {\rm cm^{3}}$ and $\omega_0 = 1.18 \, {\rm eV} = 1.794\times 10^{15} \, {\rm rad/s}$~\cite{bordag2009}. The symbol $G_{ij}^{R}$ denotes the reflection Green tensor. For an isotropic atom, we require only the $xx$, $yy$ and $zz$ components of the Green tensor, which are given by~\cite{wylie1984} 
\ba
&&G_{xx}^{R}(\rv_0,\rv_0;i\xi) = G_{yy}^{R}(\rv_0,\rv_0;i\xi) 
\\
&\!\!=\!\!& 
- \frac{\xi^2}{2c^2} \int_0^\infty \!\!\!\! dk_\parallel \, 
\frac{k_\parallel \, e^{-2\sqrt{(\xi/c)^2+k_\parallel^2} z_0}}{\sqrt{(\xi/c)^2+k_\parallel^2}} 
\nonumber\\
&&\times 
\bigg(
\frac{\sqrt{(\xi/c)^2+k_\parallel^2} - \sqrt{\varepsilon(i\xi) (\xi/c)^2+k_\parallel^2}}
{\sqrt{(\xi/c)^2+k_\parallel^2} + \sqrt{\varepsilon(i\xi) (\xi/c)^2+k_\parallel^2}}
- 
\Big(
1 + k_\parallel^2 \frac{c^2}{\xi^2}
\Big)
\nonumber\\
&&
\times 
\frac{\varepsilon(i\xi) \sqrt{(\xi/c)^2+k_\parallel^2} - \sqrt{\varepsilon(i\xi) (\xi/c)^2+k_\parallel^2}}
{\varepsilon(i\xi) \sqrt{(\xi/c)^2+k_\parallel^2} + \sqrt{\varepsilon(i\xi) (\xi/c)^2+k_\parallel^2}}
\bigg)
\nonumber
\ea
and 
\ba
&&G_{zz}^{R}(\rv_0,\rv_0;i\xi) 
\\
&\!\!=\!\!& 
\int_0^\infty \!\!\!\! dk_\parallel \, 
\frac{k_\parallel^3 e^{-2\sqrt{(\xi/c)^2+k_\parallel^2} z_0}}{\sqrt{(\xi/c)^2+k_\parallel^2}} 
\nonumber\\
&&\times 
\frac{\varepsilon(i\xi) \sqrt{(\xi/c)^2+k_\parallel^2} - \sqrt{\varepsilon(i\xi) (\xi/c)^2+k_\parallel^2}}
{\varepsilon(i\xi) \sqrt{(\xi/c)^2+k_\parallel^2} + \sqrt{\varepsilon(i\xi) (\xi/c)^2+k_\parallel^2}}. 
\nonumber
\ea
For a slab made of vitreous ${\rm SiO_2}$, we can use the Ninham-Parsegian representation for the dielectric function~\cite{bordag2009}:
\be
\varepsilon(i\xi) = 1 + \frac{C_{\rm{UV}} \, \omega_{\rm{UV}}^2}{\xi^2+\omega_{\rm{UV}}^2} 
+ \frac{C_{\rm{IR}} \, \omega_{\rm{IR}}^2}{\xi^2+\omega_{\rm{IR}}^2}, 
\label{varepsilon-SiO2}
\ee
where $C_{\rm{UV}} = 1.098$, $C_{\rm{IR}} = 1.703$, $\omega_{\rm{UV}} = 2.033\times 10^{16} \, {\rm rad/s}$, and $\omega_{\rm{IR}} = 1.88\times 10^{14} \, {\rm rad/s}$~\cite{bordag2009}. 

\section{Disorder-induced energy shift}

\subsection{A single slab with surface charge disorder} 

In this section and the following one, we study the case of an atom near a single slab. 
We first consider the case where the charge disorder only resides on the surface. 
Using Eq.~(\ref{Ev0c}), we find that the surface charge disorder-induced contribution to the energy correction described by Eq.~(\ref{energy-correction}) is given by 
\be
\delta E_n^{(S)} 
= - \frac{2 \, \alpha_{ij}^n (0)}{(\varepsilon(0)+1)^2} 
\! \int \! d^3\rv \! \int \! d^3 \rv' \, 
\frac{\overline{\delta\rho_S(\rv) \delta\rho_S(\rv')} \, n_{i}(\rv) n_{j}(\rv')}{|\rv - \rv_0|^2 |\rv' - \rv_0|^2}.  
\ee
Here, $n_i(\rv)$ is the $i$th component of the unit vector $\nv(\rv)$ which is directed from position $\rv$ towards the atom. 
Using Equation~(\ref{var-oneslab}), we obtain 
\be
\delta E_n^{(S)} 
= - \frac{2 \, \alpha_{ij}^n (0)}{(\varepsilon(0)+1)^2} 
\! \int \! d^2\rv_{\parallel} 
\frac{\sigma_S^2 n_{i}(\rv_{\parallel}) n_{j}(\rv_{\parallel})}{(r_{\parallel}^2 + z_0^2)^2}. 
\label{Enq}
\ee
In polar coordinates, $\int \! d^2\rv_{\parallel} = \int_0^{2\pi} d\varphi \int_0^\infty dr_{\parallel} \, r_{\parallel}$. 
To make progress, we need an expression for $\nv(\rv_{\parallel})$ in terms of $r_{\parallel}$, $z_0$ and $\varphi$ (which is the azimuthal angle in the xy-plane). In spherical coordinates, $\nv(\rv_{\parallel})$ is given by 
\be
\nv(\rv_{\parallel}) = - \sin\theta \cos\varphi \, {\bf e}_x 
- \sin\theta \sin\varphi \, {\bf e}_y
+ \cos\theta \, {\bf e}_z, 
\ee
where the angles $\theta$ and $\varphi$ are as shown in Fig.~\ref{fig:geometry}, and $\nv(\rv_{\parallel})$ is the unit vector directed from a given local charge disorder at $\rv_{\parallel}$ on the surface to the atom. 
Since 
$
\cos\theta = z_0/(r_{\parallel}^2 + z_0^2)^{1/2}
$ 
and 
$
\sin\theta = r_{\parallel}/(r_{\parallel}^2 + z_0^2)^{1/2}
$, 
we have that  
\ba
&&\nv(\rv_{\parallel}) 
\\
&=& n_x(\rv_{\parallel}) \, {\bf e}_x + n_y(\rv_{\parallel}) \, {\bf e}_y + n_z(\rv_{\parallel}) \, {\bf e}_z 
\nonumber\\
&=& 
- \frac{r_{\parallel} \cos\varphi}{(r_{\parallel}^2 + z_0^2)^{1/2}} \, {\bf e}_x 
- \frac{r_{\parallel} \sin\varphi}{(r_{\parallel}^2 + z_0^2)^{1/2}} \, {\bf e}_y 
+ \frac{z_0}{(r_{\parallel}^2 + z_0^2)^{1/2}} \, {\bf e}_z. 
\nonumber
\ea
After doing the angular and radial integrations, we obtain 
\be
\delta E_n^{(S)} =
- \frac{\pi \sigma_S^2 (\alpha_{xx}^n(0) + \alpha_{yy}^n(0) + 2 \alpha_{zz}^n(0))}{2 (\varepsilon(0) + 1)^2 z_0^2}. 
\label{shiftsingleslab}
\ee
The minus sign indicates that the presence of the charge disorder leads to a negative shift of the atomic energy levels. The corresponding force exerted on the atomic state is thus attractive. As $\delta E_n^{(S)}$ depends on the variance of the charge disorder density, a larger charge disorder density variance will result in a stronger attractive force. Finally, $\delta E_n^{(S)}$ decays as $1/z_0^2$, which is a slower decay compared to the nonresonant Casimir-Polder shift. 
For an isotropic atomic state, Eq.~(\ref{shiftsingleslab}) becomes 
\be
\delta E_0^{(S)} =
- \frac{2\pi \alpha_0 \sigma_S^2}{(\varepsilon(0) + 1)^2 z_0^2}. 
\label{shiftsingleslab0}
\ee
For a vitreous ${\rm{SiO_2}}$ slab surface, the static dielectric permittivity is given by $\varepsilon(0) = 3.801$. 



\subsection{A single slab with bulk charge disorder} 

Next, we derive the energy correction induced by bulk charge disorder. If we denote the $i$th impurity charge in the dielectric bulk by the symbol $\delta Q_{I}$, and let $\rv_{I}$ denote its position, the bulk charge disorder density is given by 
\be
\delta \rho_B(\rv) = \sum_{I} \delta Q_{I} \, \delta(\rv - \rv_{I}).  
\ee
By defining $\nv(\rv) = (\rv_0 - \rv)/|\rv_0 - \rv|$ and $r_{{\parallel}}^2 = x^2+y^2$, the bulk charge disorder-induced field is given by 
\be
{\bm{\mathcal{E}}}_B(\rv_0) 
= 
\int \! d^3\rv \, \frac{2 \, \delta \rho_B(\rv)}{\varepsilon(0) + 1} 
\frac{\nv(\rv)}{r_{{\parallel}}^2+(z_0-z)^2}. 
\label{EvB}
\ee
Using the above equation, we find that the bulk charge disorder-induced contribution to the energy correction described by Eq.~(\ref{energy-correction}) is given by 
\ba
\delta E_n^{(B)} 
&\!\!=\!\!& - \frac{2 \, \alpha_{ij}^n (0)}{(\varepsilon(0) + 1)^2} 
\! \int \! d^3\rv \! \int \! d^3 \rv' \, 
\frac{\overline{\delta\rho_B(\rv) \delta\rho_B(\rv')}}{r_{\parallel}^2 + (z_0-z)^2} 
\nonumber\\
&&\quad\times 
\frac{n_{i}(\rv) n_{j}(\rv')}{(r_{\parallel}^{\prime 2} + (z_0-z')^2}.   
\label{EnQ-single-slab}
\ea
Here, $n_i(\rv)$ and $n_j(\rv')$ are respectively the $i$th component of $\nv(\rv)$ and the $j$th component of $\nv(\rv')$, which are unit vectors directed from positions $\rv$ and $\rv'$ towards the atom.
After doing the disorder average using Eq.~(\ref{var-oneslab}), and performing the volume integrations, we obtain 
\be
\delta E_n^{(B)} = 
- \frac{\pi \sigma_B^2 \big( \alpha_{xx}^n(0) + \alpha_{yy}^n(0) + 2 \, \alpha_{zz}^n(0) \big)}
{2 (\varepsilon(0) + 1)^2 z_0}. 
\label{dEsingleslab}
\ee
For an isotropic atomic state, Eq.~(\ref{dEsingleslab}) becomes 
\be
\delta E_0^{(B)} = 
- \frac{2\pi \alpha_0 \sigma_B^2}
{(\varepsilon(0) + 1)^2 z_0}.
\label{dEsingleslab0}
\ee

\subsection{Two coplanar slabs with bulk charge disorder} 

 
\begin{figure}[h]
    \centering
      \includegraphics[width=0.3\textwidth]{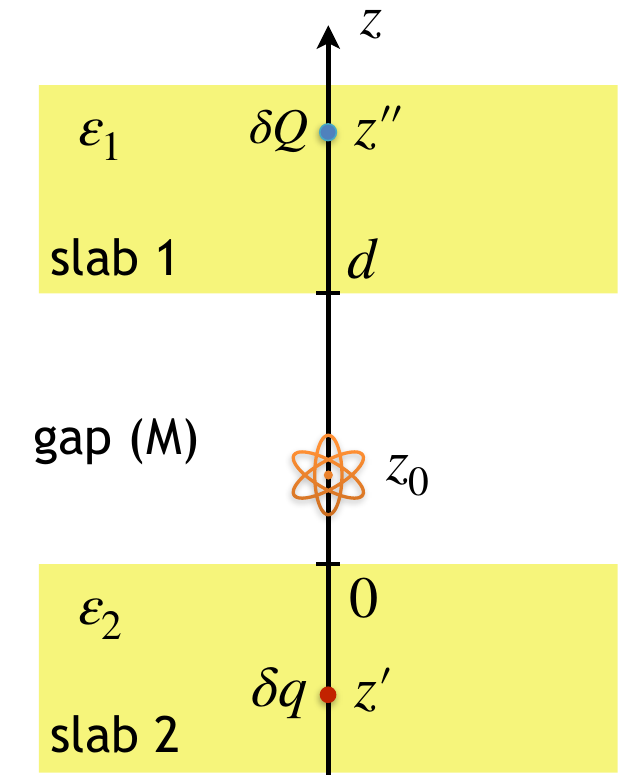}
      \caption{An atom at position $\rv_0 = z_0 \, {\bf e}_z$ between two coplanar, semi-infinite slabs with dielectric permittivities $\varepsilon_1$ and $\varepsilon_2$. The red dot at position $z' \, {\bf e}_z$ represents a given local charge disorder $\delta q$ in slab 2, whilst the blue dot at position $z'' \, {\bf e}_z$ represents a given local charge disorder $\delta Q$ in slab 1. Not shown is the rest of the charge disorder, which is homogeneously distributed throughout the bulk of each slab.} 
      \label{fig:twoslabs}
\end{figure} 

In this subsection and the next, we consider a setup which consists of two coplanar semi-infinite slabs separated by a distance $d$ in the $z$ direction, with the atom at position $\rv_0 = z_0 \, {\bf e}_z$ (see Figure~\ref{fig:twoslabs}). The top surface of the lower slab is at $z = 0$, whilst the bottom surface of the upper slab is at $z = d$. 
We consider the case where both slabs contain quenched bulk charge disorder, so the atom would experience a superposition of local electric fields generated by the totality of the bulk charge disorder. 
To find the electric field generated by a given local charge disorder, let us first consider the electric potential generated at an arbitrary point $\rv$ by a charge impurity $\delta q$ in slab 2 which has a position vector $\rv' = z' \, {\bf e}_z$ with $z' < 0$ (shown as the red dot in Figure~\ref{fig:twoslabs}). Denoting the static dielectric permittivity of the upper slab by the symbol $\varepsilon_1$ and that of the lower slab by $\varepsilon_2$, and the electric potential by the symbols $\varphi_1$, $\varphi_M$ and $\varphi_2$ for the upper slab, gap and lower slab respectively, the cylindrical symmetry about the $z$ axis enables us to write~\cite{zangwill2012} 
\ba
\label{potential}
&&\varphi(\rv_{\parallel}, z) 
\\
&=& \left\{ \begin{array}{ll}
\varphi_1(\rv_{\parallel}, z)  = \int_0^\infty \! dk \, J_0(kr_{\parallel}) C_k e^{-kz}
\quad \mbox{($z \geq d$)},
   \vspace{3mm}\\
\varphi_M(\rv_{\parallel}, z) = 
\int_0^\infty \! dk \, J_0(kr_{\parallel}) 
\big( 
A_k e^{kz} + B_k e^{-kz} 
\big)
   \vspace{3mm}\\
   \quad\qquad\qquad\qquad\qquad\qquad\qquad\qquad 
   \mbox{($0 < z < d$)},
   \vspace{3mm}\\
\varphi_2(\rv_{\parallel}, z) = 
\int_0^\infty \! dk \, J_0(kr_{\parallel}) 
\left( D_k e^{kz} + \frac{\delta q}{\varepsilon_2} e^{-k|z-z'|} \right)
   \vspace{3mm}\\
   \qquad\qquad\qquad\qquad\qquad\qquad\qquad\qquad \mbox{($z \leq 0$)}.
   \end{array}  \right.
   \nonumber
\ea
Here $J_0$ denotes a Bessel function of the first kind. The second term in the third line of the right-hand side is a rewrite of the Coulomb interaction, $(\delta q/\varepsilon_2)/(r_{\parallel}^2+(z-z')^2)^{1/2} = (\delta q/\varepsilon_2) \int_0^\infty dk \, J_0(kr_{\parallel}) \exp(-k|z-z'|)$. 
The electric potential has to satisfy the boundary conditions $\varphi_M = \varphi_2$ and $\partial \varphi_M/\partial z = \varepsilon_2 \partial \varphi_2/\partial z$ at $z = 0$, and $\varphi_M = \varphi_1$ and $\partial \varphi_M/\partial z = \varepsilon_1 \partial \varphi_1/\partial z$ at $z = d$. These boundary conditions give rise to four equations, {\it viz.} 
\ba
&&A_k + B_k - D_k = (\delta q/\varepsilon_2) e^{-k|z'|}, 
\nonumber\\
&&A_k - B_k - \varepsilon_2 D_k = - \delta q \, e^{-k|z'|}, 
\nonumber\\
&&A_k e^{kd} + B_k e^{-kd} - C_k e^{-kd} = 0, 
\nonumber\\
&&A_k e^{kd} - B_k e^{-kd} + \varepsilon_1 C_k e^{-kd} = 0. 
\ea
We solve the equations for $A_k$, $B_k$, $C_k$ and $D_k$:   
\ba
&&A_k = - \frac{2 \delta q \, e^{-k(|z'|+2d)}}{\varepsilon_2+1} \frac{\Delta_1}{1-\Delta_1\Delta_2e^{-2kd}}, 
\nonumber\\
&&B_k = \frac{2 \delta q \, e^{-k|z'|}}{\varepsilon_2+1} \frac{1}{1-\Delta_1\Delta_2e^{-2kd}}, 
\nonumber\\
&&C_k = \frac{4 \delta q \, e^{-k|z'|}}{(\varepsilon_1+1)(\varepsilon_2+1)}
\frac{1}{1-\Delta_1\Delta_2e^{-2kd}}, 
\nonumber\\
&&D_k = \frac{\delta q \, e^{-k|z'|}}{\varepsilon_2}
\frac{\Delta_2 - \Delta_1 e^{-2kd}}
{1-\Delta_1\Delta_2e^{-2kd}},
\ea
where the dielectric discontinuity factors are given by 
\be
\Delta_1 = \frac{\varepsilon_1 - 1}{\varepsilon_1 + 1}, \quad
\Delta_2 = \frac{\varepsilon_2 - 1}{\varepsilon_2 + 1}. 
\ee
The denominator factor $(1-\Delta_1\Delta_2e^{-2kd})^{-1}$ accounts for the contribution of the (infinitely many) image charges. 
To see this, let us expand the denominator factor in powers of $\Delta_1\Delta_2e^{-2kd}$:  
\ba
A_k e^{kz} &=& - \frac{2\delta q}{\varepsilon_2 + 1} \sum_{J=0}^{\infty} \Delta_1^{J+1} \Delta_2^J e^{k(z+z'-2(J+1)d)}, 
\nonumber \\
B_k e^{-kz} &=& \frac{2\delta q}{\varepsilon_2 + 1} \sum_{J=0}^{\infty} \Delta_1^J \Delta_2^J e^{-k(z-z'+2Jd)}. 
\ea
We can thus express $\varphi_M$ in Equation~(\ref{potential}) as 
\ba
&&\varphi_M(\rv_{\parallel}, z) 
\nonumber\\
&=& 
\frac{2\delta q}{\varepsilon_2 + 1} 
\sum_{J=0}^{\infty} 
\Bigg(
- \frac{\Delta_1^{J+1} \Delta_2^J}{\sqrt{r_{\parallel}^2 + (z+z'-2(J+1)d)^2}} 
\nonumber\\
&&\quad+
\frac{\Delta_1^{J} \Delta_2^J}{\sqrt{r_{\parallel}^2 + (z-z'+2Jd)^2}} 
\Bigg). 
\ea
From the above expression, it follows that the potential at a position $\rv$ inside the gap generated by an impurity charge $\delta q_{I}$ located at $\rv_{I}$ in the lower slab is given by 
\ba
&&\varphi_{I}(\rv) = 
\frac{2\delta q_{I}}{\varepsilon_2 + 1} 
\sum_{J=0}^{\infty} 
\nonumber\\
&& 
\Bigg(
- \frac{\Delta_1^{J+1} \Delta_2^J}{\sqrt{(x - x_{I})^2 + (y-y_{I})^2 + (z+z_{I}-2(J+1)d)^2}} 
\nonumber\\
&&\quad+
\frac{\Delta_1^{J} \Delta_2^J}{\sqrt{(x - x_{I})^2 + (y-y_{I})^2 + (z-z_{I}+2Jd)^2}} 
\Bigg).
\label{varphii}
\ea
\begin{figure}[h]
    \centering
      \includegraphics[width=0.48\textwidth]{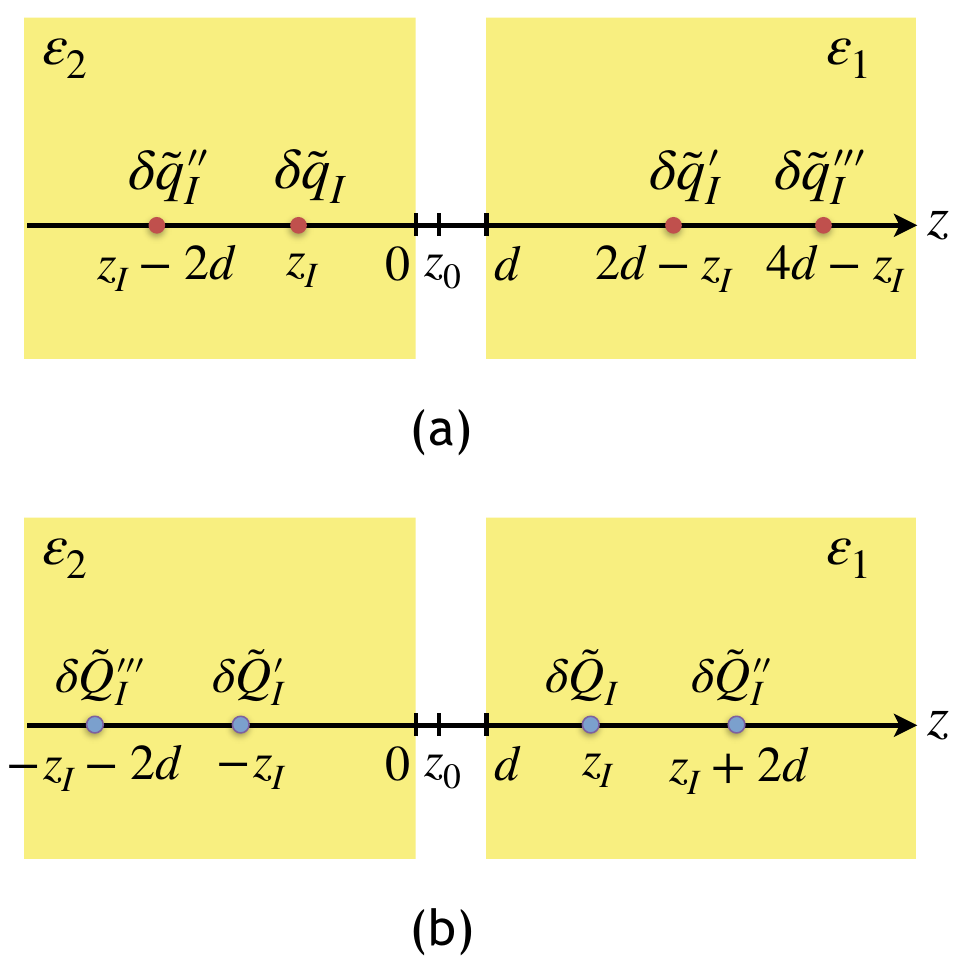}
      \caption{(a)~Interpretation of the terms in the series for $\varphi_{I}$ (Eq.~(\ref{varphii})): $\delta \tilde{q}_{I} \equiv 2\delta q_{I}/(\varepsilon_2+1)$ is the image charge of the impurity charge $\delta q_{I}$ positioned at $z=z_{I}$ ($z_{I} < 0$) ``seen" by the atom at $z = z_0$ if only slab 2 (with dielectric permittivity $\varepsilon_2$) is present. Adding slab 1 (with dielectric permittivity $\varepsilon_1$) gives rise to an image charge of $\delta \tilde{q}_{I}$, {\it i.e.}, $\delta \tilde{q}_{I'} = - \Delta_1 \delta \tilde{q}_{I}$ at $z=2d-z_{I}$. In turn, $\delta \tilde{q}_{I'}$ gives rise to another image charge $\delta \tilde{q}_{I''} = \Delta_2\Delta_1 \delta \tilde{q}_{I}$ at $z=z_{I}-2d$, and this image charge gives rise to a further image charge $\delta \tilde{q}_{I'''} = -\Delta_2\Delta_1^2 \delta \tilde{q}_{I}$ at $z=4d-z_{I}$, and so on. 
      (b)~Interpretation of the terms in the series for $\psi_{I}$ (Eq.~(\ref{psii})): 
$\delta \tilde{Q}_{I} \equiv 2\delta Q_{I}/(\varepsilon_1+1)$ is the image charge of the impurity charge $\delta Q_{I}$ positioned at $z=z_{I}$ ($z_{I} > 0$) ``seen" by the atom at $z = z_0$ if only slab 1 (with dielectric permittivity $\varepsilon_1$) is present.      
Adding slab 2 (with dielectric permittivity $\varepsilon_2$) gives rise to an image charge of $\delta \tilde{Q}_{I}$, {\it i.e.}, $\delta \tilde{Q}_{I'} = - \Delta_2 \delta \tilde{Q}_{I}$ at $z=-z_{I}$.
      In turn, $\delta \tilde{Q}_{I'}$ gives rise to another image charge $\delta \tilde{Q}_{I''} = \Delta_1\Delta_2 \delta \tilde{Q}_{I}$ at $z=z_{I}+2d$, and this image charge gives rise to a further image charge $\delta \tilde{Q}_{I'''} = -\Delta_1\Delta_2^2 \delta \tilde{Q}_{I}$ at $z=-z_{I}-2d$, and so on.} 
      \label{fig:image}
\end{figure} 
If we write out the terms of the series explicitly, we see that they correspond to the contribution from image charges. This situation is depicted by Figure~\ref{fig:image}(a) (for the case where $x_{I}=y_{I}=0$). 
The electric field corresponding to $\varphi_{I}(\rv)$ is given by 
\ba
&&{\bm{\mathcal{E}}}_{I}(\rv) 
\\
&\!\!=\!\!& 
\frac{2\delta q_{I}}{\varepsilon_2 + 1} 
\sum_{J=0}^{\infty} 
\Bigg(
\frac{\Delta_1^{J} \Delta_2^J \, \nv(\rv_{I})}{(x - x_{I})^2 + (y-y_{I})^2 + (z - z_{I}+2Jd)^2} 
\nonumber\\
&&\quad -
\frac{\Delta_1^{J+1} \Delta_2^J \, \mv(\rv_{I})}{(x - x_{I})^2 + (y-y_{I})^2 + (z + z_{I}-2(J+1)d)^2} 
\Bigg), 
\nonumber
\ea
where 
\ba
\nv(\rv_{I}) &\!\!=\!\!& 
\frac{(x-x_{I}) \, {\bf e}_x + (y-y_{I}) \, {\bf e}_y + (z-z_{I}+2Jd) \, {\bf e}_z}
{\sqrt{(x - x_{I})^2 + (y-y_{I})^2 + (z-z_{I}+2Jd)^2}}, 
\nonumber\\
\mv(\rv_{I}) &\!\!=\!\!& 
\frac{(x-x_{I}) \, {\bf e}_x + (y-y_{I}) \, {\bf e}_y + (z + z_{I}-2(J+1)d) \, {\bf e}_z}
{\sqrt{(x - x_{I})^2 + (y-y_{I})^2 + (z + z_{I}-2(J+1)d)^2}}. 
\nonumber\\
\ea
At the atom's position $\rv_0 = z_0\,{\bf e}_z$, ${\bm{\mathcal{E}}}_{I}$ becomes 
\ba
{\bm{\mathcal{E}}}_{I}(\rv_0) &\!\!=\!\!& 
\frac{2\delta q_{I}}{\varepsilon_2 + 1} 
\sum_{J=0}^{\infty} 
\Bigg(
\frac{\Delta_1^{J} \Delta_2^J \, \nv(\rv_{I})}{r_{{I},{\parallel}}^2 + (z_0 - z_{I}+2Jd)^2} 
\nonumber\\
&&\quad -
\frac{\Delta_1^{J+1} \Delta_2^J \, \mv(\rv_{I})}{r_{{I},{\parallel}}^2 + (z_0 + z_{I}-2(J+1)d)^2} 
\Bigg), 
\label{Ei}
\ea
where
\ba
\nv(\rv_{I}) &\!\!=\!\!& 
\frac{-x_{I} \, {\bf e}_x -y_i \, {\bf e}_y + (z_0-z_{I}+2Jd) \, {\bf e}_z}
{\sqrt{r_{{I},{\parallel}}^2 + (z_0-z_{I}+2Jd)^2}}, 
\nonumber\\
\mv(\rv_{I}) &\!\!=\!\!& 
\frac{-x_{I} \, {\bf e}_x -y_{I} \, {\bf e}_y + (z_0 + z_{I}-2(J+1)d) \, {\bf e}_z}
{\sqrt{r_{{I},{\parallel}}^2 + (z_0 + z_{I}-2(J+1)d)^2}}. 
\nonumber\\
\ea
By introducing the local bulk impurity charge density in slab 2, {\it viz.}, 
\be
\delta \rho_2(\rv) = \sum_{I} \delta q_{I} \, \delta(\rv - \rv_{I}), 
\ee
and summing the electric field in Equation~(\ref{Ei}) over all impurity charges in slab 2, the total field at $\rv_0$ exerted by the impurity charges in slab 2 is given by 
\ba
{\bm{\mathcal{E}}}(\rv_0) &=& 
\frac{2}{\varepsilon_2 + 1} 
\int_{-\infty}^{0} \!\!\!\! dz 
\int \! d^2\rv_{\parallel} 
\delta \rho_2(\rv)
\nonumber\\
&&\times 
\sum_{J=0}^{\infty} 
\Bigg(
\frac{\Delta_1^{J} \Delta_2^J \, \nv(\rv)}{r_{{\parallel}}^2 + (z_0 - z + 2Jd)^2} 
\nonumber\\
&&\quad
-
\frac{\Delta_1^{J+1} \Delta_2^J \, \mv(\rv)}{r_{{\parallel}}^2 + (z_0 + z - 2(J+1)d)^2} 
\Bigg), 
\ea
where
\begin{subequations}
\ba
\nv(\rv) &\!\!=\!\!& 
\frac{-x \, {\bf e}_x -y \, {\bf e}_y + (z_0-z+2Jd) \, {\bf e}_z}
{\sqrt{r_{{\parallel}}^2 + (z_0-z+2Jd)^2}}, 
\nonumber\\
\mv(\rv) &\!\!=\!\!& 
\frac{-x \, {\bf e}_x -y \, {\bf e}_y + (z_0 + z-2(J+1)d) \, {\bf e}_z}
{\sqrt{r_{{\parallel}}^2 + (z_0 + z-2(J+1)d)^2}}. 
\nonumber\\
\ea
\end{subequations}
The above calculation pertains to the field induced at $\rv_0$ by impurity charges in slab 2.  

The calculation of the field induced at $\rv_0$ by impurity charges in slab 1 is analogous (see App.~\ref{app:slab1charges}), and yields 
\ba
\label{F-slab1}
&&{\bm{\mathcal{F}}}(\rv_0) = \frac{2}{\varepsilon_1 + 1} 
\int_{d}^{\infty} \!\!\!\! dz \int \! d^2\rv_{\parallel} 
\delta \rho_1(\rv)
\\
&&\times
\sum_{J=0}^{\infty} 
\Bigg(
\frac{\Delta_1^{J} \Delta_2^J \, \Nv(\rv)}{r_{{\parallel}}^2 + (z_0 - z - 2Jd)^2} 
-
\frac{\Delta_1^{J} \Delta_2^{J+1} \, \Mv(\rv)}{r_{{\parallel}}^2 + (z_0 + z + 2Jd)^2} 
\Bigg), 
\nonumber
\ea
where $\delta \rho_1(\rv)$ denotes the local bulk impurity charge density in slab 1, {\it viz.}, 
\be
\delta \rho_1(\rv) = \sum_{I} \delta Q_{I} \, \delta(\rv - \rv_{I}), 
\label{deltarho1}
\ee
\begin{subequations}
and the direction vectors $\Nv(\rv)$ and $\Mv(\rv)$ are defined by 
\ba
\Nv(\rv) &=& 
\frac{-x \, {\bf e}_x -y \, {\bf e}_y + (z_0 - z - 2Jd) \, {\bf e}_z}
{\sqrt{r_{{\parallel}}^2 + (z_0 - z - 2Jd)^2}}, 
\label{N-slab1}
\\
\Mv(\rv) &=& 
\frac{-x \, {\bf e}_x -y \, {\bf e}_y + (z_0 + z +2Jd) \, {\bf e}_z}
{\sqrt{r_{{\parallel}}^2 + (z_0 + z + 2Jd)^2}}. 
\label{M-slab1}
\ea
\end{subequations}
The bulk charge disorder-averaged contribution to the energy correction described by Eq.~(\ref{energy-correction}) is now given by 
\ba
&&\delta E_n^{(BB)}(\rv_0) 
\label{EnQ-twoslabs}\\
&\!\!=\!\!& 
- \frac{1}{2} \alpha_{ij}^n (0) 
\overline{(\mathcal{E}_{{i}}(\rv_0)+\mathcal{F}_{{i}}(\rv_0))
(\mathcal{E}_{{j}}(\rv_0)+\mathcal{F}_{{j}}(\rv_0))}. 
\nonumber
\ea
As before, we assume the charge disorder in each slab is Gaussian distributed with zero mean, {\it i.e.}, $\overline{\delta\rho_1} = \overline{\delta\rho_2} = 0$, and the variance of the local charge disorder per unit volume is $\sigma_{B1}^2$ in slab 1 and $\sigma_{B2}^2$ in slab 2, with the disorder being spatially uncorrelated:  
\ba
&&\overline{\delta\rho_1(\rv) \delta\rho_2(\rv')} = 0, \,\,
\overline{\delta\rho_1(\rv) \delta\rho_1(\rv')} 
= \sigma_{B1}^2 \, \delta(\rv - \rv'), 
\nonumber\\
&&\overline{\delta\rho_2(\rv) \delta\rho_2(\rv')} 
= \sigma_{B2}^2 \, \delta(\rv - \rv').
\label{var-twoslabs}
\ea
By performing the disorder average in Eq.~(\ref{EnQ-twoslabs}) and integrating over $\rv_{\parallel}$, we obtain (see Appendix~\ref{app:EnBB} for details)
\be
\delta E_n^{(BB)} = \delta E_{n,1}^{(BB)} + \delta E_{n,2}^{(BB)}, 
\label{EnBB}
\ee
where
\ba
\delta E_{n,1}^{(BB)} 
&\!\!=\!\!& 
- \frac{\pi \sigma_{B1}^2 (\alpha_{xx}^n (0) + \alpha_{yy}^n (0) + 2\alpha_{zz}^n (0))}{2(\varepsilon_1 + 1)^2} 
\nonumber\\
&&\times
\Bigg(
\frac{1}{d} \Phi\left(\Delta_1\Delta_2, 1, 1 - \frac{z_0}{d}\right)
\nonumber\\
&&
+ \frac{\Delta_1\Delta_2}{2d-z_0} {}_2F_1\left( 2, 2-\frac{z_0}{d}; 3-\frac{z_0}{d}; \Delta_1\Delta_2 \right)
\nonumber\\
&&
+  
\frac{\Delta_2^{2}}{d} \Phi\left(\Delta_1\Delta_2, 1, 1 + \frac{z_0}{d}\right)
\nonumber\\
&&
+ \frac{\Delta_1\Delta_2^3}{2d+z_0} {}_2F_1\left( 2, 2+\frac{z_0}{d}; 3+\frac{z_0}{d}; \Delta_1\Delta_2 \right)
\Bigg)
\nonumber\\
&&
+ \frac{\pi \sigma_{B1}^2 (\alpha_{xx}^n (0) + \alpha_{yy}^n (0) - 2\alpha_{zz}^n (0)) \Delta_2}{(\varepsilon_1 + 1)^2 (1-\Delta_1\Delta_2) d} 
\nonumber
\ea
and 
\ba
\delta E_{n,2}^{(BB)} 
&\!\!=\!\!& 
- \frac{\pi \sigma_{B2}^2 (\alpha_{xx}^n (0) + \alpha_{yy}^n (0) + 2 \alpha_{zz}^n (0))}{2(\varepsilon_2 + 1)^2} 
\nonumber\\
&&\times
\Bigg(
\frac{1}{d} \Phi\left(\Delta_1\Delta_2, 1, \frac{z_0}{d}\right) 
\nonumber\\
&&
+ \frac{\Delta_1\Delta_2}{d+z_0} {}_2F_1\left( 2, 1+\frac{z_0}{d}; 2+\frac{z_0}{d}; \Delta_1\Delta_2 \right)
\nonumber\\
&&
+ \frac{\Delta_1^{2}}{d} \Phi\left(\Delta_1\Delta_2, 1, 2 - \frac{z_0}{d}\right)
\nonumber\\
&&
+ \frac{\Delta_1^3\Delta_2}{3d-z_0} {}_2F_1\left( 2, 3-\frac{z_0}{d}; 4-\frac{z_0}{d}; \Delta_1\Delta_2 \right) 
\Bigg)
\nonumber\\
&&
+ \frac{\pi \sigma_{B2}^2 (\alpha_{xx}^n (0) + \alpha_{yy}^n (0) - 2 \alpha_{zz}^n (0)) \Delta_1}{(\varepsilon_2 + 1)^2 (1-\Delta_1\Delta_2) d}.   
\nonumber
\ea
Here, ${}_2F_1(a,b;c;z)$ is the Gauss hypergeometric function and $\Phi(z,s,a)$ is the Lerch transcendent, defined by $\Phi(z,s,a) \equiv \sum_{J=0}^{\infty} z^J/(a+J)^s$~\cite{nist}. From the definition of the Lerch transcendent, we see that $\Phi(z,s,a)$ diverges as $a \to 0$. This implies that as $z_0/d \to 0$, the prefactor $\frac{1}{d} \Phi\left(\Delta_1\Delta_2, 1, \frac{z_0}{d}\right)$ in the equation above diverges and becomes dominant over the other terms. 
As we show in App.~\ref{app:EnBB}, as $d \to \infty$, we recover the result for an atom near a single charge-disordered slab (Eq.~(\ref{dEsingleslab})). 

\subsection{Two coplanar slabs with surface charge disorder} 

The boundary value problem for surface charge disorder for two coplanar slabs is analogous to the one considered previously, with the modifications that the impurity charges $\delta q_{I}$ are located just below the surface of slab 2, and the impurity charges $\delta Q_{I}$ are located just above the surface of slab 1. 
The charge disorder densities $\delta\rho_1(\rv)$ and $\delta\rho_2(\rv)$ for slabs 1 and 2 are now given by  
\ba
&&\delta\rho_1(\rv) = \sum_{I} Q_{I} \delta(\rv_{\parallel} - \rv_{{I},{\parallel}})\delta(z - d+),
\nonumber\\
&&\delta\rho_2(\rv) = \sum_{I} q_{I} \delta(\rv_{\parallel} - \rv_{{I},{\parallel}}) \delta(z - 0-). 
\ea
Correspondingly, the fields ${\bm{\mathcal{E}}}_S(\rv_0)$ and ${\bm{\mathcal{F}}}_S(\rv_0)$ at the atom's position $\rv_0 = z_0\,{\bf e}_z$ generated respectively by the images of surface impurity charges on slab 2 and the images of surface impurity charges on slab 1 are given by 
\begin{subequations}
\ba
&&{\bm{\mathcal{E}}}_S(\rv_0) 
=
\frac{2}{\varepsilon_2 + 1} 
\int_{-\infty}^{0} \!\!\!\! dz 
\int \! d^2\rv_{\parallel} 
\delta \rho_2(\rv)
\\
&& 
\sum_{J=0}^{\infty} 
\Bigg(
\frac{\Delta_1^{J} \Delta_2^J \, \nv(\rv_{\parallel})}{r_{{\parallel}}^2 + (z_0 + 2Jd)^2} 
-
\frac{\Delta_1^{J+1} \Delta_2^J \, \mv(\rv_{\parallel})}{r_{{\parallel}}^2 + (z_0 - 2(J+1)d)^2} 
\Bigg), 
\nonumber\\
&&{\bm{\mathcal{F}}}_S(\rv_0) 
=
\frac{2}{\varepsilon_1 + 1} 
\int_{d}^{\infty} \!\!\!\! dz 
\int \! d^2\rv_{\parallel} 
\delta \rho_1(\rv)
\\
&&
\sum_{J=0}^{\infty} 
\Bigg(
\frac{\Delta_1^{J} \Delta_2^J \, \Nv(\rv_{\parallel})}{r_{{\parallel}}^2 + (z_0 - (2J+1) d)^2} 
-
\frac{\Delta_1^{J} \Delta_2^{J+1} \, \Mv(\rv_{\parallel})}{r_{{\parallel}}^2 + (z_0 + (2J+1) d)^2} 
\Bigg). 
\nonumber  
\ea
\end{subequations}
Here, the direction vectors $\nv(\rv_{\parallel})$, $\mv(\rv_{\parallel})$, $\Nv(\rv_{\parallel})$ and $\Mv(\rv_{\parallel})$ are given by 
\ba
\nv(\rv_{\parallel}) &=& 
\frac{-x \, {\bf e}_x -y \, {\bf e}_y + (z_0 + 2Jd) \, {\bf e}_z}
{\sqrt{r_{{\parallel}}^2 + (z_0 + 2Jd)^2}}, 
\nonumber\\
\mv(\rv_{\parallel}) &=& 
\frac{-x \, {\bf e}_x -y \, {\bf e}_y + (z_0 - 2(J+1)d) \, {\bf e}_z}
{\sqrt{r_{{\parallel}}^2 + (z_0 - 2(J+1)d)^2}}, 
\nonumber\\
\Nv(\rv_{\parallel}) &=& 
\frac{-x \, {\bf e}_x -y \, {\bf e}_y + (z_0 - (2J+1) d) \, {\bf e}_z}
{\sqrt{r_{{\parallel}}^2 + (z_0 - (2J+1) d)^2}}, 
\nonumber\\
\Mv(\rv_{\parallel}) &=& 
\frac{-x \, {\bf e}_x -y \, {\bf e}_y + (z_0 + (2J+1) d) \, {\bf e}_z}
{\sqrt{r_{{\parallel}}^2 + (z_0 + (2J+1) d)^2}}.
\ea
As before, we assume the surface charge disorder is Gaussian distributed with zero mean, {\it i.e.}, $\overline{\delta\rho_1} = \overline{\delta\rho_2} = 0$, and the variance of the local surface charge disorder per unit area is $\sigma_{S1}^2$ in slab 1 and $\sigma_{S2}^2$ in slab 2, with the disorder being spatially uncorrelated:  
\ba
&&\overline{\delta\rho_1(\rv) \delta\rho_1(\rv')} 
= \sigma_{S1}^2 \, \delta(\rv_{\parallel} - \rv_{\parallel}') 
\delta(z - d+) \delta(z' - d+), 
\nonumber\\
&&\overline{\delta\rho_2(\rv) \delta\rho_2(\rv')} 
= \sigma_{S2}^2 \, \delta(\rv_{\parallel} - \rv_{\parallel}')
\delta(z - 0-) \delta(z' - 0-),  
\nonumber\\
&&\overline{\delta\rho_1(\rv) \delta\rho_2(\rv')} = 0.  
\label{var-twosurfaces}
\ea
The surface charge disorder-averaged contribution to the energy correction described by Eq.~(\ref{energy-correction}) is given by 
\ba
&&\delta E_n^{(SS)}(\rv_0) 
\label{EnQ-twosurfaces}\\
&\!\!=\!\!& 
- \frac{1}{2} \alpha_{ab}^n (0) 
\overline{(\mathcal{E}_{Sa}(\rv_0)+\mathcal{F}_{Sa}(\rv_0))
(\mathcal{E}_{Sb}(\rv_0)+\mathcal{F}_{Sb}(\rv_0))}. 
\nonumber
\ea
After performing the disorder average and the integration over $\rv_{\parallel}$, we obtain (see Appendix~\ref{app:EnSS} for details) 
\ba
&&\delta E_n^{(SS)} 
=
- \frac{\pi \sigma_{S1}^2 (\alpha_{xx}^n (0) + \alpha_{yy}^n (0) + 2\alpha_{zz}^n (0))}{2(\varepsilon_1 + 1)^2} 
\nonumber\\
&&\times
\Bigg(
\frac{1}{d^2} \Phi\left(\Delta_1\Delta_2,1,1-\frac{z_0}{d}\right) 
+ \frac{z_0}{d^3} \Phi\left(\Delta_1\Delta_2,2,1-\frac{z_0}{d}\right) 
\nonumber\\
&&
+ \frac{\Delta_2^2 }{d^2} \Phi\left(\Delta_1\Delta_2,1,1+\frac{z_0}{d}\right) 
- \frac{\Delta_2^2  z_0}{d^3} \Phi\left(\Delta_1\Delta_2,2,1+\frac{z_0}{d}\right) 
\Bigg)
\nonumber\\
&&- \frac{\pi \sigma_{S2}^2 (\alpha_{xx}^n (0) + \alpha_{yy}^n (0) + 2 \alpha_{zz}^n (0))}{2(\varepsilon_2 + 1)^2} 
\nonumber\\
&&\times
\Bigg(
\frac{1}{d^2} \Phi\left(\Delta_1\Delta_2,1,\frac{z_0}{d}\right) 
+ \frac{d-z_0}{d^3} \Phi\left(\Delta_1\Delta_2,2,\frac{z_0}{d}\right)    
\nonumber\\
&&\quad+ 
\frac{\Delta_1^2}{d^2} \Phi\left(\Delta_1\Delta_2,1,2-\frac{z_0}{d}\right) 
\nonumber\\
&&
\quad+\frac{\Delta_1^2 (z_0-d)}{d^3} \Phi\left(\Delta_1\Delta_2,2,2-\frac{z_0}{d}\right) 
\Bigg)
\nonumber\\
&&
- \pi (\alpha_{xx}^n (0) + \alpha_{yy}^n (0) - 2\alpha_{zz}^n (0)) \ln (1-\Delta_1\Delta_2)
\nonumber\\
&&\quad\times
\left(
\frac{\sigma_{S1}^2}{\Delta_1 (\varepsilon_1 + 1)^2}
+
\frac{\sigma_{S2}^2}{\Delta_2 (\varepsilon_2 + 1)^2}
\right)
 \frac{1}{d^2} .   
 \label{EnSS}
\ea
In App.~\ref{app:EnSSproof}, we show that the above result leads to Eq.~(\ref{shiftsingleslab}) as $d \to \infty$, which is the surface charge disorder-induced energy level shift in an atom near a single slab.





\section{Results and Discussion}

\begin{figure}[h]
    \centering
      \includegraphics[width=0.5\textwidth]{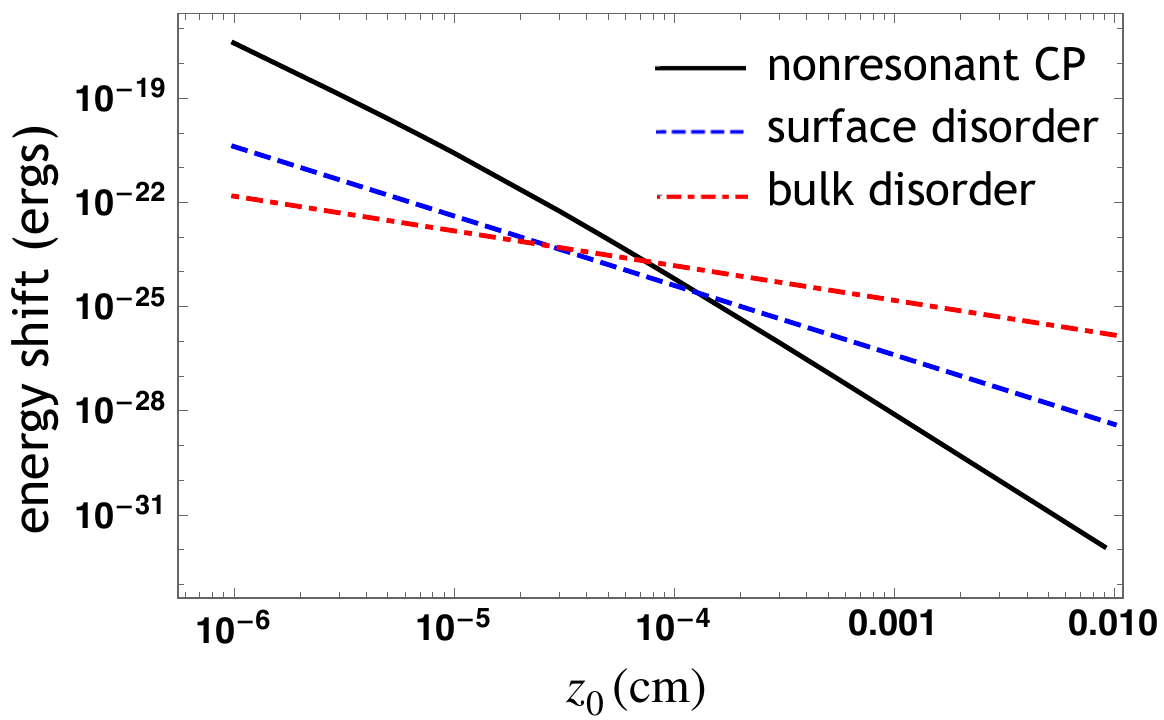}
      \caption{Log-log plots of the magnitude of the nonresonant Casimir-Polder energy shift (black), surface charge disorder-induced energy shift (blue dashed) and bulk charge disorder-induced energy shift (red dot-dashed) as functions of the atom-surface separation $z_0$, for a  helium atom in the $n=2$ triplet state next to a vitreous ${\rm{SiO_2}}$ slab surface, with polarizability and dielectric functions given by Eqs.~(\ref{alpha-He*}) and (\ref{varepsilon-SiO2}). We set $\sigma_B^2 = 1.168\times 10^{-5} \, {\rm esu^2/cm^3}$ and $\sigma_S^2 = 3.146 \times 10^{-10} \, {\rm esu^2/cm^2}$ for the bulk and surface charge disorder variances per unit area.} 
      \label{fig:plots-compare1}
\end{figure} 
\begin{figure}[h]
    \centering
      \includegraphics[width=0.5\textwidth]{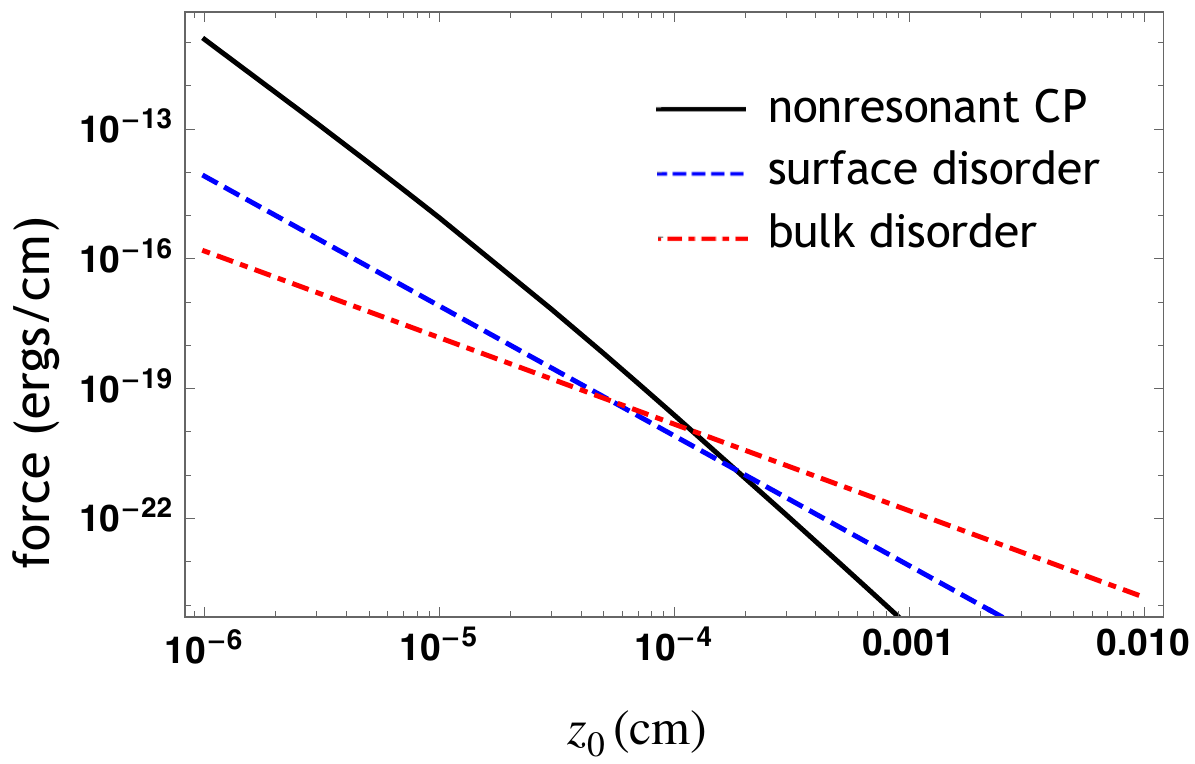}
      \caption{Log-log plots of the magnitude of the nonresonant Casimir-Polder force (black), surface charge disorder-induced force (blue dashed) and bulk charge disorder-induced force (red dot-dashed) as functions of the atom-surface separation $z_0$, for a  helium atom in the $n=2$ triplet state next to a vitreous ${\rm{SiO_2}}$ slab surface, with polarizability and dielectric functions given by Eqs.~(\ref{alpha-He*}) and (\ref{varepsilon-SiO2}). We set $\sigma_B^2 = 1.168\times 10^{-5} \, {\rm esu^2/cm^3}$ and $\sigma_S^2 = 3.146 \times 10^{-10} \, {\rm esu^2/cm^2}$ for the bulk and surface charge disorder variances per unit area.} 
      \label{fig:plots-compare2}
\end{figure} 
To illustrate the applications of the formulas derived in the previous section, let us consider the example of a  helium atom in the $n=2$ triplet state and assume the slab material is made of vitreous ${\rm{SiO_2}}$. The corresponding static polarizability is $\alpha_0 = 315.63 \, {\rm a.u.} = 4.678\times 10^{-23} \, {\rm cm^{3}}$, and the static dielectric permittivity is $\varepsilon = 3.801$~\cite{bordag2009}. 

\

In Fig.~\ref{fig:plots-compare1}, we plot the behaviors of the surface disorder-induced energy shift (blue dashed line) and bulk disorder-induced energy shifts (red dot-dashed line) described by Eqs.~(\ref{shiftsingleslab0}) and (\ref{dEsingleslab0}) for the case of the atom next to a single slab surface, with the bulk charge disorder variance per unit volume and the surface charge disorder variance per unit area set (as in Ref.~\cite{naji2010}) to the values $\sigma_B^2 = 1.168\times 10^{-5} \, {\rm esu^2/cm^3}$ and $\sigma_S^2 = 3.146 \times 10^{-10} \, {\rm esu^2/cm^2}$ respectively~\cite{footnote-esu}. 
In the same figure, we also plot the behavior of the nonresonant Casimir-Polder (CP) energy shift of the  helium atom in the $n=2$ triplet state for comparison. 
We see that the nonresonant CP shift experiences a crossover in its power-law decay behavior as the separation increases from smaller than $10^{-5}$ cm (near-field region) to greater than $10^{-4}$ cm (far-field region). 
This crossover lengthscale is set by the dominant transition wavelength of the atom, which in this case corresponds to $1.05 \times 10^{-4}$ cm.
There is another crossover scale of the order of $10^{-4}$ cm, below which the surface  disorder-induced shift (which decays as $z_0^{-2}$) and bulk disorder-induced shift (which decays as $z_0^{-1}$) are smaller than the nonresonant CP shift, and above which the disorder-induced shifts dominate over the nonresonant CP shift. 

\

In Fig.~\ref{fig:plots-compare2}, we compare the magnitudes of the nonresonant CP force, surface disorder-induced force and bulk disorder-induced force for the same material parameters, also finding crossovers occurring at similar orders of magnitude of the separation distance.  
Because the disorder-induced shifts and forces depend linearly on the charge disorder variance per unit area, an increase in the charge disorder leads to an upward shift in the magnitude of the disorder-induced shifts and forces. This also implies that the crossover at which the disorder-induced force starts dominating over the nonresonant CP force will occur at smaller separations. Conversely, such a crossover will occur at larger separations if the charge disorder is decreased. 
From these results, we expect that if a Casimir-Polder force measurement is done on an atom near a slab having an unknown charge disorder variance in the far-field region, and the force behavior is found to decay as $z_0^{-2}$ ($z_0^{-3}$) instead of $z_0^{-5}$ as expected for the far-field region of the Casimir-Polder interaction, we may interpret the decay behavior as being due to the presence of bulk (surface) charge impurities, and the value of the bulk (surface) charge disorder variance per unit volume (area) may also be extrapolated from the gradient of the graph.  

\

\begin{figure}[h]
    \centering
      \includegraphics[width=0.48\textwidth]{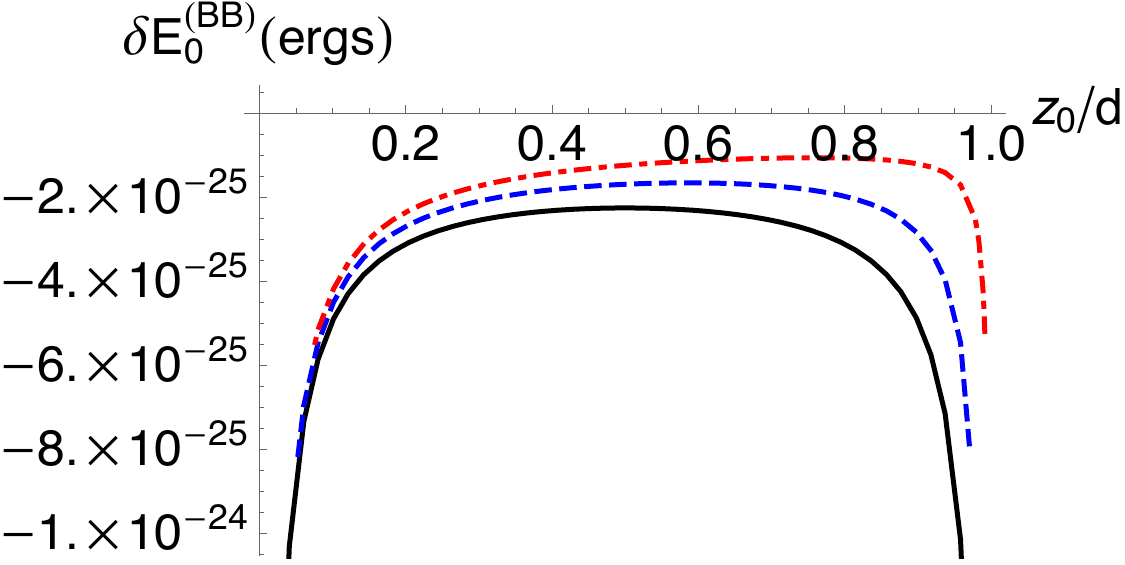}
      \caption{Behavior of the bulk charge disorder-induced energy shift $\delta {E}_0^{(BB)}$ for a  helium atom in the $n=2$ triplet state confined between coplanar and infinitely thick vitreous ${{\rm SiO_2}}$ slab surfaces for $\sigma_{B1}^2 = \sigma_{B2}^2$ (black), $\sigma_{B1}^2 = 0.5 \sigma_{B2}^2$ (blue dashed), and $\sigma_{B1}^2 = 0.1 \sigma_{B2}^2$ (red, dot-dashed). Here, $z_0$ is the distance of the atom from the surface of slab 2, and we set $d = 4 \times 10^{-3}$ cm as the separation between the slab surfaces.} 
      \label{fig:E0BB}
\end{figure} 
\begin{figure}[h]
    \centering
      \includegraphics[width=0.48\textwidth]{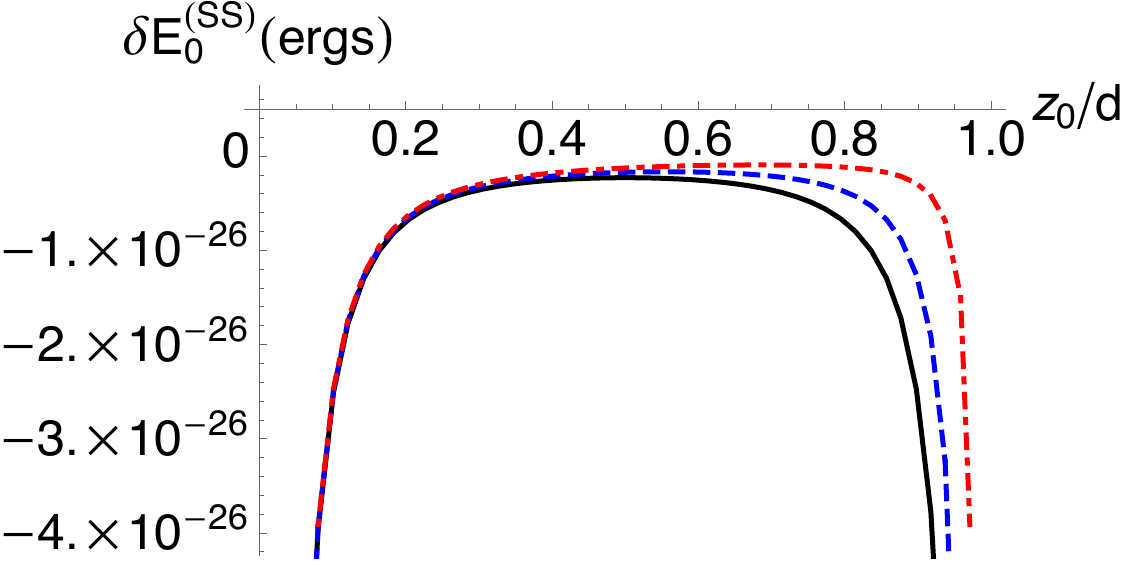}
      \caption{Behavior of the surface charge disorder-induced energy shift $\delta {E}_0^{(SS)}$ for a  helium atom in the $n=2$ triplet state confined between coplanar and infinitely thick vitreous ${{\rm SiO_2}}$ slab surfaces for $\sigma_{S1}^2 = \sigma_{S2}^2$ (black), $\sigma_{S1}^2 = 0.5 \sigma_{S2}^2$ (blue dashed), and $\sigma_{S1}^2 = 0.1 \sigma_{S2}^2$ (red, dot-dashed). Here, $z_0$ is the distance of the atom from the surface of slab 2, and we set $d = 4 \times 10^{-3}$ cm as the separation between the slab surfaces.} 
      \label{fig:E0SS}
\end{figure} 
In Fig.~\ref{fig:E0BB}, we show the behavior of the bulk charge disorder-induced energy shift $\delta {E}_0^{(BB)}$ as a function of $z_0$ (the distance between the atom and the surface of slab 2) for the atom confined between two coplanar and infinitely thick slabs, whose surfaces are $4 \times 10^{-3}$ cm apart. We consider the following three different cases: (i)~$\sigma_{B1}^2 = \sigma_{B2}^2$, (ii)~$\sigma_{B1}^2 = 0.5 \sigma_{B2}^2$, and (iii)~$\sigma_{B1}^2 = 0.1 \sigma_{B2}^2$. 
We see that for the case where both slabs have the same bulk charge disorder variance per unit volume, the gradient of the disorder-induced energy shift is zero at the exact center ($z_0 = d/2$), where the atom is equally attracted to both surfaces. 
On the other hand, as the bulk charge disorder variance of slab 1 decreases relative to that of slab 2, the position of zero gradient shifts towards slab 2. As the bulk charge disorder-induced attractive force due to a slab is proportional to the slab's bulk charge disorder variance and becomes weaker the further away the atom is from the slab, the attractive force that slab 1 (on the right) exerts on the atom only cancels that exerted by slab 2 (on the left) at a position closer to slab 1.  

\

In Fig.~\ref{fig:E0SS}, we display the behavior of the surface charge disorder-induced energy shift $\delta {E}_0^{(SS)}$ as a function of $z_0$ for the atom confined between two coplanar and infinitely thick slabs, whose surfaces are held at a distance of $4 \times 10^{-3}$ cm apart. Similar to the previous example, we consider the three cases: (i)~$\sigma_{S1}^2 = \sigma_{S2}^2$, (ii)~$\sigma_{S1}^2 = 0.5 \sigma_{S2}^2$, and (iii)~$\sigma_{S1}^2 = 0.1 \sigma_{S2}^2$. 
The gradient of the surface disorder-induced energy shift is zero at the exact center ($z_0 = d/2$) if both slabs have the same surface disorder variance. 
Analogous to what we have seen for the previous example, as the surface charge disorder variance of slab 1 is reduced relative to that of slab 2, the zero gradient position shifts towards slab 2.

\section{Summary and Conclusion} 

In this paper, we have investigated the atom-surface interaction behavior induced by surface and bulk charge disorders present in the dielectric material, for the case where the atom-surface separation is larger than the charge disorder correlation length. In particular, we have studied the disorder-induced interaction behavior of an atom next to a single planar semi-infinite dielectric slab as well as the disorder-induced energy level shifts for an atom confined between two coplanar and semi-infinite slabs. As an example, we take for the atom a  helium atom in the $n=2$ triplet state, and we assume the slab material is vitreous ${{\rm SiO_2}}$. 
For the case where only a single slab is present, we have found that the surface (bulk) charge disorder leads to a downward energy shift which decays as $z_0^{-2}$ ($z_0^{-1}$) (where $z_0$ is the perpendicular distance of the atom from the surface), with a corresponding force that is attractive and decays as $z_0^{-3}$ ($z_0^{-2}$). We also found that the disorder-induced force will dominate the nonresonant Casimir-Polder force at sufficiently large distances, and the distance at which this occurs is larger (smaller) for smaller (larger) values of the charge disorder variance per unit area. For the particular case where the bulk charge disorder variance per unit volume is $1.168\times 10^{-5} \, {\rm esu^2/cm^3}$ and the surface charge disorder variance per unit area is $\sigma_S^2 = 3.146 \times 10^{-10} \, {\rm esu^2/cm^2}$, such a crossover occurs at a separation of the order of $10^{-4}$ cm. 

\

Lastly, we have studied the behavior of the charge disorder-induced energy shift for an atom confined in the vacuum gap between two coplanar and semi-infinite slabs made of the same dielectric material. For the case of surface disorder as well as bulk disorder, we found that if the charge disorder variance is equal on both slabs, the atom experiences a net zero disorder-induced force when it is positioned right in the middle of the gap, whereas if the charge disorder variance on one of the slabs is smaller than the other one, the position of the net zero charge disorder-induced force (which is an unstable equilibrium position) shifts towards the slab with the smaller charge disorder variance. 

\

Although we did not discuss in this paper, it is possible to extend our investigations to the case of charge disorder with a nonzero correlation length, which would be relevant for atom-surface separations of the order of a few nanometers. 
One can also use the same formalism to study the charge disorder-induced energy level shifts in an excited and anisotropically polarized atom near a dielectric slab. 
One can also use a similar approach to study the case of a Rydberg atom between two dielectric slabs with quenched monopolar charge disorder. For this case, one would need to account for multipolar contributions higher than the dipolar one.

\section{acknowledgments} 

The author thanks David Wilkowski for a constructive discussion, and is grateful to Derek Frydel for hosting him at the Universidad T\'{e}cnica Federico Santa Mar\'{i}a, where part of the work was carried out. 
He thanks the two anonymous referees for their constructive feedback. The author dedicates this paper to the memory of Rudolf Podgornik, who was a wonderfully kind and inspiring mentor and friend.

\begin{widetext} 

\appendix

\section{Derivation of ${\bm{\mathcal{F}}}(\rv_0)$ (Eq.~(\ref{F-slab1}))}
\label{app:slab1charges}
 
For the case of a single charge impurity $\delta Q$ in slab 1, with position $\rv' = z'' \, {\bf e}_z$ where now $z'' > d$ (see Figure~\ref{fig:twoslabs}), the potential exerted at $\rv$ by this charge is 
\ba
\label{potential1}
\psi(\rv_{\parallel}, z) 
&=& \left\{ \begin{array}{ll}
\psi_1(\rv_{\parallel}, z)  = \int_0^\infty \! dk \, J_0(kr_{\parallel}) 
\left( C_k e^{-kz} + \frac{\delta Q}{\varepsilon_1} e^{-k|z-z''|} \right)
\quad   \mbox{($z \geq d$)},
   \vspace{3mm}\\
\psi_M(\rv_{\parallel}, z) = 
\int_0^\infty \! dk \, J_0(kr_{\parallel}) 
\big( 
A_k e^{kz} + B_k e^{-kz} 
\big)
\qquad\qquad 
\mbox{($0 < z < d$)},
   \vspace{3mm}\\
\psi_2(\rv_{\parallel}, z) = 
\int_0^\infty \! dk \, J_0(kr_{\parallel}) D_k e^{kz}  
  \qquad\qquad\qquad\qquad\qquad 
  \mbox{($z \leq 0$)}.
   \end{array}  \right.
\ea
Applying the boundary conditions $\psi_M = \psi_2$ and $\partial \psi_M/\partial z = \varepsilon_2 \partial \psi_2/\partial z$ at $z = 0$, and $\psi_M = \psi_1$ and $\partial \psi_M/\partial z = \varepsilon_1 \partial \psi_1/\partial z$ at $z = d$ leads to  
\ba
&&A_k + B_k - D_k = 0, 
\nonumber\\
&&A_k - B_k - \varepsilon_2 D_k = 0, 
\nonumber\\
&&A_k e^{kd} + B_k e^{-kd} - C_k e^{-kd} = (\delta Q/\varepsilon_1) e^{-k(z'' - d)}, 
\nonumber\\
&&A_k e^{kd} - B_k e^{-kd} + \varepsilon_1 C_k e^{-kd} = \delta Q \, e^{-k(z''-d)}. 
\ea
Solving for $A_k$, $B_k$, $C_k$ and $D_k$, we obtain 
\ba
&&A_k = \frac{2 \delta Q \, e^{-k z''}}{\varepsilon_1+1} 
\frac{1}{1-\Delta_1\Delta_2e^{-2kd}}, 
\nonumber\\
&&B_k = - \frac{2 \delta Q \, e^{-k z''}}{\varepsilon_1+1} 
\frac{\Delta_2}{1-\Delta_1\Delta_2e^{-2kd}}, 
\nonumber\\
&&C_k = \frac{\delta Q \, e^{-k(z''-2d)}}{\varepsilon_1}
\frac{\Delta_1 - \Delta_2 e^{-2kd}}
{1-\Delta_1\Delta_2e^{-2kd}},
\nonumber\\
&&D_k = \frac{4 \delta Q \, e^{-k z''}}{(\varepsilon_1+1)(\varepsilon_2+1)}
\frac{1}{1-\Delta_1\Delta_2e^{-2kd}}, 
\ea
Using the series expansion for $(1-\Delta_1\Delta_2e^{-2kd})^{-1}$ in powers of $\Delta_1\Delta_2e^{-2kd}$ allows us to write 
\ba
&&A_k e^{kz} = \frac{2\delta Q}{\varepsilon_1 + 1} 
\sum_{J=0}^{\infty} 
\Delta_1^{J} \Delta_2^J 
e^{ - k(z'' - z + 2Jd)}, 
\nonumber\\
&&B_k e^{-kz} = - \frac{2\delta Q}{\varepsilon_1 + 1} 
\sum_{J=0}^{\infty} \Delta_1^J \Delta_2^{J+1} 
e^{-k(z'' + z + 2Jd)}, 
\ea
and thus the potential $\psi_M$ exerted at a point $\rv$ inside the gap by $\delta Q$ is given by 
\ba
\psi_M(\rv_{\parallel}, z) =
\frac{2\delta Q}{\varepsilon_1 + 1} 
\sum_{J=0}^{\infty} 
\Bigg(
\frac{\Delta_1^{J} \Delta_2^J}{\sqrt{r_{\parallel}^2 + (z-z''-2Jd)^2}} 
-
\frac{\Delta_1^{J} \Delta_2^{J+1}}{\sqrt{r_{\parallel}^2 + (z+z''+2Jd)^2}} 
\Bigg). 
\ea
From the expression above, it follows that the potential at a position $\rv$ inside the gap generated by a charge impurity $\delta Q_{I}$ located at $\rv_{I}$ in the upper slab is given by 
\ba
\label{psii}
\psi_{I}(\rv) 
=
\frac{2\delta Q_{I}}{\varepsilon_1 + 1} 
\sum_{J=0}^{\infty} 
\Bigg(
\frac{\Delta_1^{J} \Delta_2^J}{\sqrt{(x-x_{I})^2 + (y-y_{I})^2 + (z-z_{I}-2Jd)^2}} 
-
\frac{\Delta_1^{J} \Delta_2^{J+1}}{\sqrt{(x-x_{I})^2 + (y-y_{I})^2 + (z+z_{I}+2Jd)^2}} 
\Bigg). 
\ea
Analogous to Eq.~(\ref{varphii}), the terms in the series also represent image charge contributions, which we illustrate in Figure~\ref{fig:image}(b). 
The corresponding electric field is 
\ba
{\bm{\mathcal{F}}}_{I}(\rv) 
=
\frac{2\delta Q_{I}}{\varepsilon_1 + 1} 
\sum_{J=0}^{\infty} 
\Bigg(
\frac{\Delta_1^{J} \Delta_2^J \, \Nv(\rv_{I})}{(x - x_{I})^2 + (y-y_{I})^2 + (z - z_{I}-2Jd)^2} 
-
\frac{\Delta_1^{J} \Delta_2^{J+1} \, \Mv(\rv_{I})}{(x - x_{I})^2 + (y-y_{I})^2 + (z + z_{I} + 2Jd)^2} 
\Bigg), 
\ea
where 
\ba
\Nv(\rv_{I}) &=& 
\frac{(x-x_{I}) \, {\bf e}_x + (y-y_{I}) \, {\bf e}_y + (z-z_{I}-2Jd) \, {\bf e}_z}
{\sqrt{(x - x_{I})^2 + (y-y_{I})^2 + (z-z_{I}-2Jd)^2}}, 
\nonumber\\
\Mv(\rv_{I}) &=& 
\frac{(x-x_{I}) \, {\bf e}_x + (y-y_{I}) \, {\bf e}_y + (z + z_{I} + 2Jd) \, {\bf e}_z}
{\sqrt{(x - x_{I})^2 + (y-y_{I})^2 + (z + z_{I} + 2Jd)^2}}. 
\ea
At the atom's position $\rv_0 = z_0\,{\bf e}_z$, ${\bm{\mathcal{F}}}_i$ becomes 
\ba
{\bm{\mathcal{F}}}_{I}(\rv_0) &=& 
\frac{2\delta Q_{I}}{\varepsilon_1 + 1} 
\sum_{J=0}^{\infty} 
\Bigg(
\frac{\Delta_1^{J} \Delta_2^J \, \Nv(\rv_{I})}{r_{{I},{\parallel}}^2 + (z_0 - z_{I} - 2Jd)^2} 
-
\frac{\Delta_1^{J} \Delta_2^{J+1} \, \Mv(\rv_{I})}{r_{{I},{\parallel}}^2 + (z_0 + z_{I} + 2Jd)^2} 
\Bigg), 
\label{Fi}
\ea
where
$
\Nv(\rv_{I}) =
\frac{- x_{I} \, {\bf e}_x - y_{I} \, {\bf e}_y + (z_0 - z_{I} - 2Jd) \, {\bf e}_z}
{\sqrt{r_{{I},{\parallel}}^2 + (z_0 - z_{I} - 2Jd)^2}}$ and 
$\Mv(\rv_{I}) =
\frac{- x_{I} \, {\bf e}_x  - y_{I} \, {\bf e}_y + (z_0 + z_{I} + 2Jd) \, {\bf e}_z}
{\sqrt{r_{{I},{\parallel}}^2 + (z_0 + z_{I} + 2Jd)^2}}$. 
By using the local charge density defined by Eq.~(\ref{deltarho1}) and summing the electric field in Equation~(\ref{Fi}) over the impurity charges in slab 1, we obtain the total field at $\rv_0$ exerted by the impurity charges in slab 1, Eq.~(\ref{F-slab1}). 

\section{Derivation of $\delta E_n^{(BB)}$ (Eq.~(\ref{EnBB}))}
\label{app:EnBB}

Performing the disorder average in Equation~(\ref{EnQ-twoslabs}) yields 
\ba
\delta E_n^{(BB)} 
&=& 
- \frac{2\sigma_{B1}^2 \alpha_{{ij}}^n (0)}{(\varepsilon_1 + 1)^2} 
\sum_{J=0}^\infty \sum_{K=0}^\infty
\int_{d}^{\infty} \!\!\!\! dz \int \! d^2\rv_{\parallel} 
\left(
\frac{\Delta_1^{J} \Delta_2^J \, N_{i}(\rv)}{r_{{\parallel}}^2 + (z_0 - z - 2Jd)^2} 
-
\frac{\Delta_1^{J} \Delta_2^{J+1} \, M_{i}(\rv)}{r_{{\parallel}}^2 + (z_0 + z + 2Jd)^2} 
\right) 
\nonumber\\
&&\quad\times
\left(
\frac{\Delta_1^{K} \Delta_2^K \, N_{j}(\rv)}{r_{{\parallel}}^2 + (z_0 - z - 2Kd)^2} 
-
\frac{\Delta_1^{K} \Delta_2^{K+1} \, M_{j}(\rv)}{r_{{\parallel}}^2 + (z_0 + z + 2Kd)^2} 
\right)
\nonumber\\
&&- \frac{2\sigma_{B2}^2 \alpha_{{ij}}^n (0)}{(\varepsilon_2 + 1)^2} 
\sum_{J=0}^\infty \sum_{K=0}^\infty
\int_{-\infty}^{0} \!\!\!\! dz \int \! d^2\rv_{\parallel} 
\left(
\frac{\Delta_1^{J} \Delta_2^J \, n_{i}(\rv)}{r_{{\parallel}}^2 + (z_0 - z + 2Jd)^2} 
-
\frac{\Delta_1^{J+1} \Delta_2^J \, m_{i}(\rv)}{r_{{\parallel}}^2 + (z_0 + z - 2(J+1)d)^2} 
\right)
\nonumber\\
&&\quad\times
\left(
\frac{\Delta_1^{K} \Delta_2^K \, n_{j}(\rv)}{r_{{\parallel}}^2 + (z_0 - z + 2Kd)^2} 
-
\frac{\Delta_1^{K+1} \Delta_2^K \, m_{j}(\rv)}{r_{{\parallel}}^2 + (z_0 + z - 2(K+1)d)^2} 
\right).   
\label{energy}
\ea
In terms of cylindrical coordinates, the integral measure $\int d^2\rv_{\parallel} = \int_0^\infty dr_{\parallel} r_{\parallel} \int_0^{2\pi} d\varphi$, and 
\begin{subequations}
\ba
\Nv(\rv) &=& 
\frac{r_{\parallel} \cos\varphi \, {\bf e}_x + r_{\parallel} \sin\varphi  \, {\bf e}_y + (z_0 - z - 2Jd) \, {\bf e}_z}
{\sqrt{r_{{\parallel}}^2 + (z_0 - z - 2Jd)^2}} \qquad\quad\,\, (z > 0), 
\\
\Mv(\rv) &=& 
\frac{r_{\parallel} \cos\varphi \, {\bf e}_x + r_{\parallel} \sin\varphi \, {\bf e}_y + (z_0 + z +2Jd) \, {\bf e}_z}
{\sqrt{r_{{\parallel}}^2 + (z_0 + z + 2Jd)^2}} \qquad\quad\,\, (z > 0), 
\\
\nv(\rv) &=& 
\frac{r_{\parallel} \cos\varphi \, {\bf e}_x + r_{\parallel} \sin\varphi \, {\bf e}_y + (z_0-z+2Jd) \, {\bf e}_z}
{\sqrt{r_{{\parallel}}^2 + (z_0-z+2Jd)^2}} \qquad\quad\,\, (z < 0), 
\\
\mv(\rv) &=& 
\frac{r_{\parallel} \cos\varphi \, {\bf e}_x + r_{\parallel} \sin\varphi \, {\bf e}_y + (z_0 + z-2(J+1)d) \, {\bf e}_z}
{\sqrt{r_{{\parallel}}^2 + (z_0 + z-2(J+1)d)^2}} \quad (z < 0).
\ea
\end{subequations}
As $\int_0^{2\pi} d\varphi \, \sin\varphi \cos\varphi = \int_0^{2\pi} d\varphi \, \sin\varphi  = \int_0^{2\pi} d\varphi \, \cos\varphi = 0$, the off-diagonal terms in Equation~(\ref{energy}) vanish when we integrate over $\varphi$. 
Since $\int_0^{2\pi} d\varphi \, \cos^2\varphi = \int_0^{2\pi} d\varphi \, \sin^2\varphi = \pi$, we obtain  
\ba
\delta E_n^{(BB)} 
&\!\!=\!\!& 
- \frac{2\pi \sigma_{B1}^2 (\alpha_{xx}^n (0) + \alpha_{yy}^n (0))}{(\varepsilon_1 + 1)^2} 
\sum_{J=0}^\infty \sum_{K=0}^\infty
\int_{d}^{\infty} \!\!\!\! dz 
\int_0^\infty \!\!\!\! dr_{\parallel} \, r_{\parallel}^3
\left(
\frac{\Delta_1^{J} \Delta_2^J}{(r_{{\parallel}}^2 + (z - z_0 + 2Jd)^2)^{3/2}} 
-
\frac{\Delta_1^{J} \Delta_2^{J+1}}{(r_{{\parallel}}^2 + (z + z_0 + 2Jd)^2)^{3/2}} 
\right) 
\nonumber\\
&&\quad\times
\left(
\frac{\Delta_1^{K} \Delta_2^K}{(r_{{\parallel}}^2 + (z - z_0 + 2Kd)^2)^{3/2}} 
-
\frac{\Delta_1^{K} \Delta_2^{K+1}}{(r_{{\parallel}}^2 + (z + z_0 + 2Kd)^2)^{3/2}} 
\right)
\nonumber\\
&&- \frac{4\pi \sigma_{B1}^2 \alpha_{zz}^n (0)}{(\varepsilon_1 + 1)^2} 
\sum_{J=0}^\infty \sum_{K=0}^\infty
\int_{d}^{\infty} \!\!\!\! dz 
\int_0^\infty \!\!\!\! dr_{\parallel} \, r_{\parallel}
\left(
\frac{\Delta_1^{J} \Delta_2^J (z_0 - z - 2Jd)}{(r_{{\parallel}}^2 + (z - z_0 + 2Jd)^2)^{3/2}} 
-
\frac{\Delta_1^{J} \Delta_2^{J+1} (z_0 + z + 2Jd)}{(r_{{\parallel}}^2 + (z + z_0 + 2Jd)^2)^{3/2}} 
\right) 
\nonumber\\
&&\quad\times
\left(
\frac{\Delta_1^{K} \Delta_2^K (z_0 - z - 2Kd)}{(r_{{\parallel}}^2 + (z - z_0 + 2Kd)^2)^{3/2}} 
-
\frac{\Delta_1^{K} \Delta_2^{K+1} (z_0 + z + 2Kd)}{(r_{{\parallel}}^2 + (z + z_0 + 2Kd)^2)^{3/2}} 
\right)
\nonumber\\
&&- \frac{2\pi \sigma_{B2}^2 (\alpha_{xx}^n (0) + \alpha_{yy}^n (0))}{(\varepsilon_2 + 1)^2} 
\sum_{J=0}^\infty \sum_{K=0}^\infty
\int_{-\infty}^{0} \!\!\!\! dz 
\int_0^\infty \!\!\!\! dr_{\parallel} \, r_{\parallel}^3
\Bigg(
\frac{\Delta_1^{J} \Delta_2^J}{(r_{{\parallel}}^2 + (z - z_0 - 2Jd)^2)^{3/2}} 
\nonumber\\
&&-
\frac{\Delta_1^{J+1} \Delta_2^J}{(r_{{\parallel}}^2 + (z + z_0 - 2(J+1)d)^2)^{3/2}} 
\Bigg)
\left(
\frac{\Delta_1^{K} \Delta_2^K}{(r_{{\parallel}}^2 + (z - z_0 - 2Kd)^2)^{3/2}} 
-
\frac{\Delta_1^{K+1} \Delta_2^K}{(r_{{\parallel}}^2 + (z + z_0 - 2(K+1)d)^2)^{3/2}} 
\right)
\nonumber\\
&&- \frac{4\pi \sigma_{B2}^2 \alpha_{zz}^n (0)}{(\varepsilon_2 + 1)^2} 
\sum_{J=0}^\infty \sum_{K=0}^\infty
\int_{-\infty}^{0} \!\!\!\! dz 
\int_0^\infty \!\!\!\! dr_{\parallel} \, r_{\parallel}
\left(
\frac{\Delta_1^{J} \Delta_2^J (z_0 - z + 2Jd)}{(r_{{\parallel}}^2 + (z - z_0 - 2Jd)^2)^{3/2}} 
-
\frac{\Delta_1^{J+1} \Delta_2^J (z_0 + z - 2(J+1)d)}{(r_{{\parallel}}^2 + (z + z_0 - 2(J+1)d)^2)^{3/2}} 
\right)
\nonumber\\
&&\quad\times
\left(
\frac{\Delta_1^{K} \Delta_2^K (z_0 - z + 2Kd)}{(r_{{\parallel}}^2 + (z - z_0 - 2Kd)^2)^{3/2}} 
-
\frac{\Delta_1^{K+1} \Delta_2^K (z_0 + z - 2(K+1)d)}{(r_{{\parallel}}^2 + (z + z_0 - 2(K+1)d)^2)^{3/2}} 
\right).   
\nonumber
\ea
The integrals over $r_{\parallel}$ can be performed by making use of the following identities: 
\begin{subequations}
\ba
&&\int_0^\infty \!\!\!\! dr_{\parallel} \frac{r_{\parallel}^3}{(r_{\parallel}^2 + A^2)^{3/2} (r_{\parallel}^2 + B^2)^{3/2}} = \frac{1}{(|A|+|B|)^2}, 
\\
&&\int_0^\infty \!\!\!\! dr_{\parallel} \frac{r_{\parallel}}{(r_{\parallel}^2 + A^2)^{3/2} (r_{\parallel}^2 + B^2)^{3/2}} = \frac{1}{|A||B|(|A|+|B|)^2}. 
\ea
\end{subequations}
We obtain 
\ba
\delta E_n^{(BB)} 
&\!\!=\!\!& 
- \frac{\pi \sigma_{B1}^2 (\alpha_{xx}^n (0) + \alpha_{yy}^n (0))}{2(\varepsilon_1 + 1)^2} 
\sum_{J=0}^\infty \sum_{K=0}^\infty
\int_{d}^{\infty} \!\!\!\! dz 
\left(
\frac{\Delta_1^{J+K} \Delta_2^{J+K}}{(z-z_0+Jd+Kd)^2} 
- \frac{2\,\Delta_1^{J+K} \Delta_2^{J+K+1}}{(z+Jd+Kd)^2} 
+ \frac{\Delta_1^{J+K} \Delta_2^{J+K+2}}{(z+z_0+Jd+Kd)^2} 
\right)
\nonumber\\
&&- \frac{\pi \sigma_{B1}^2 \alpha_{zz}^n (0)}{(\varepsilon_1 + 1)^2} 
\sum_{J=0}^\infty \sum_{K=0}^\infty
\int_{d}^{\infty} \!\!\!\! dz 
\left(
\frac{\Delta_1^{J+K} \Delta_2^{J+K}}{(z-z_0+Jd+Kd)^2} 
+ \frac{2\,\Delta_1^{J+K} \Delta_2^{J+K+1}}{(z+Jd+Kd)^2} 
+ \frac{\Delta_1^{J+K} \Delta_2^{J+K+2}}{(z+z_0+Jd+Kd)^2} 
\right)
\nonumber\\
&&- \frac{\pi \sigma_{B2}^2 (\alpha_{xx}^n (0) + \alpha_{yy}^n (0))}{2(\varepsilon_2 + 1)^2} 
\sum_{J=0}^\infty \sum_{K=0}^\infty
\int_{-\infty}^{0} \!\!\!\! dz 
\Bigg(
\frac{\Delta_1^{J+K} \Delta_2^{J+K}}{(z-z_0-Jd-Kd)^2} 
- \frac{\Delta_1^{J+K+1} \Delta_2^{J+K}}{(z-(J+1)d-Kd)^2} 
\nonumber\\
&&\quad
- \frac{\Delta_1^{J+K+1} \Delta_2^{J+K}}{(z-Jd-(K+1)d)^2} 
+ \frac{\Delta_1^{J+K+2} \Delta_2^{J+K}}{(z+z_0-(J+1)d-(K+1)d)^2} 
\Bigg)
\nonumber\\
&&- \frac{\pi \sigma_{B2}^2 \alpha_{zz}^n (0)}{(\varepsilon_2 + 1)^2} 
\sum_{J=0}^\infty \sum_{K=0}^\infty
\int_{-\infty}^{0} \!\!\!\! dz 
\Bigg(
\frac{\Delta_1^{J+K} \Delta_2^{J+K}}{(z-z_0-Jd-Kd)^2} 
+ \frac{\Delta_1^{J+K+1} \Delta_2^{J+K}}{(z-(J+1)d-Kd)^2} 
\nonumber\\
&&\quad
+ \frac{\Delta_1^{J+K+1} \Delta_2^{J+K}}{(z-Jd-(K+1)d)^2} 
+ \frac{\Delta_1^{J+K+2} \Delta_2^{J+K}}{(z+z_0-(J+1)d-(K+1)d)^2} 
\Bigg).   
\nonumber
\ea
After performing the integral over $z$, we obtain 
\ba
\delta E_n^{(BB)} 
&\!\!=\!\!& 
- \frac{\pi \sigma_{B1}^2 (\alpha_{xx}^n (0) + \alpha_{yy}^n (0))}{2(\varepsilon_1 + 1)^2} 
\sum_{J=0}^\infty \sum_{K=0}^\infty
\left(
\frac{(\Delta_1\Delta_2)^{J+K}}{(J+K+1)d-z_0} 
- 2 \Delta_2 \frac{(\Delta_1\Delta_2)^{J+K}}{(J+K+1)d} 
+ \Delta_2^{2} \frac{(\Delta_1\Delta_2)^{J+K}}{(J+K+1)d+z_0} 
\right)
\nonumber\\
&&- \frac{\pi \sigma_{B1}^2 \alpha_{zz}^n (0)}{(\varepsilon_1 + 1)^2} 
\sum_{J=0}^\infty \sum_{K=0}^\infty
\left(
\frac{(\Delta_1\Delta_2)^{J+K}}{(J+K+1)d-z_0} 
+ 2 \Delta_2 \frac{(\Delta_1\Delta_2)^{J+K}}{(J+K+1)d} 
+ \Delta_2^2 \frac{(\Delta_1\Delta_2)^{J+K}}{(J+K+1)d+z_0} 
\right)
\nonumber\\
&&- \frac{\pi \sigma_{B2}^2 (\alpha_{xx}^n (0) + \alpha_{yy}^n (0))}{2(\varepsilon_2 + 1)^2} 
\sum_{J=0}^\infty \sum_{K=0}^\infty
\Bigg(
\frac{(\Delta_1\Delta_2)^{J+K}}{(J+K)d+z_0}
- 2 \Delta_1 \frac{(\Delta_1\Delta_2)^{J+K}}{(J+K+1)d} 
+ \Delta_1^2 \frac{(\Delta_1\Delta_2)^{J+K}}{(J+K+2)d-z_0} 
\Bigg)
\nonumber\\
&&- \frac{\pi \sigma_{B2}^2 \alpha_{zz}^n (0)}{(\varepsilon_2 + 1)^2} 
\sum_{J=0}^\infty \sum_{K=0}^\infty
\Bigg(
\frac{(\Delta_1\Delta_2)^{J+K}}{(J+K)d+z_0}  
+ 2 \Delta_1 \frac{(\Delta_1\Delta_2)^{J+K}}{(J+K+1)d} 
+ \Delta_1^2 \frac{(\Delta_1\Delta_2)^{J+K}}{(J+K+2)d-z_0} 
\Bigg).   
\ea
We check that we recover the result for an atom above a single charge-disordered slab: if we let $d \to \infty$, all the terms on the right-hand side vanish, except the $J=K=0$ contribution to the two terms which are $\frac{(\Delta_1\Delta_2)^{J+K}}{(J+K)d+z_0}$, which give 
\be
\delta E_n^{(BB)} \to - \frac{\pi \sigma_{B2}^2 (\alpha_{xx}^n (0) + \alpha_{yy}^n (0) + 2 \alpha_{zz}^n (0))}
{2 (\varepsilon_2 + 1)^2 z_0}.   
\ee 
This is the same result as Equation~(\ref{dEsingleslab}), with $\sigma_{B2}^2$ replaced by $\sigma_B^2$ and $\varepsilon_2$ replaced by $\varepsilon$. 
Finally, doing the double summation over $J$ and $K$ leads to Eq.~(\ref{EnBB}). 

\section{Derivation of $\delta E_n^{(SS)}$ (Eq.~(\ref{EnSS}))}
\label{app:EnSS}

In this Appendix, we provide the calculational steps for obtaining Eq.~(\ref{EnSS}) from Eq.~(\ref{EnQ-twosurfaces}). 

Performing the disorder average in Eq.~(\ref{EnQ-twosurfaces}) yields 
\ba
\delta E_n^{(SS)} 
&=&
- \frac{2\sigma_{S1}^2 \alpha_{{ij}}^n (0)}{(\varepsilon_1 + 1)^2} 
\sum_{J=0}^\infty \sum_{K=0}^\infty
\int \! d^2\rv_{\parallel} 
\left(
\frac{\Delta_1^{J} \Delta_2^J \, N_{i}(\rv_{\parallel})}{r_{{\parallel}}^2 + (z_0 - (2J+1) d)^2} 
-
\frac{\Delta_1^{J} \Delta_2^{J+1} \, M_{i}(\rv_{\parallel})}{r_{{\parallel}}^2 + (z_0 + (2J+1) d)^2} 
\right) 
\nonumber\\
&&\times
\left(
\frac{\Delta_1^{K} \Delta_2^K \, N_{j}(\rv_{\parallel})}{r_{{\parallel}}^2 + (z_0 - (2K+1) d)^2} 
-
\frac{\Delta_1^{K} \Delta_2^{K+1} \, M_{j}(\rv_{\parallel})}{r_{{\parallel}}^2 + (z_0 + (2K+1) d)^2} 
\right)
\nonumber\\
&&- \frac{2\sigma_{S2}^2 \alpha_{{ij}}^n (0)}{(\varepsilon_2 + 1)^2} 
\sum_{J=0}^\infty \sum_{K=0}^\infty
\int \! d^2\rv_{\parallel} 
\left(
\frac{\Delta_1^{J} \Delta_2^J \, n_{i}(\rv_{\parallel})}{r_{{\parallel}}^2 + (z_0  + 2Jd)^2} 
-
\frac{\Delta_1^{J+1} \Delta_2^J \, m_{i}(\rv_{\parallel})}{r_{{\parallel}}^2 + (z_0 - 2(J+1)d)^2} 
\right)
\nonumber\\
&&\times
\left(
\frac{\Delta_1^{K} \Delta_2^K \, n_{j}(\rv_{\parallel})}{r_{{\parallel}}^2 + (z_0 + 2Kd)^2} 
-
\frac{\Delta_1^{K+1} \Delta_2^K \, m_{j}(\rv_{\parallel})}{r_{{\parallel}}^2 + (z_0 - 2(K+1)d)^2} 
\right).   
\nonumber\\
\label{energy1}
\ea
As before, we switch to polar coordinates and integrate over the azimuthal angle. Only diagonal terms survive this integration, and we obtain 
\ba
\delta E_n^{(SS)} 
&=& 
- \frac{\pi \sigma_{S1}^2 (\alpha_{xx}^n (0) + \alpha_{yy}^n (0))}{2(\varepsilon_1 + 1)^2} 
\sum_{J=0}^\infty \sum_{K=0}^\infty
\left(
\frac{\Delta_1^{J+K} \Delta_2^{J+K}}{((J+K+1)d - z_0)^2} 
- \frac{2\,\Delta_1^{J+K} \Delta_2^{J+K+1}}{(J+K+1)^2 d^2} 
+ \frac{\Delta_1^{J+K} \Delta_2^{J+K+2}}{((J+K+1)d + z_0)^2} 
\right)
\nonumber\\
&&- \frac{\pi \sigma_{S1}^2 \alpha_{zz}^n (0)}{(\varepsilon_1 + 1)^2} 
\sum_{J=0}^\infty \sum_{K=0}^\infty
\left(
\frac{\Delta_1^{J+K} \Delta_2^{J+K}}{((J+K+1)d - z_0)^2} 
+ \frac{2\,\Delta_1^{J+K} \Delta_2^{J+K+1}}{(J+K+1)^2 d^2} 
+ \frac{\Delta_1^{J+K} \Delta_2^{J+K+2}}{((J+K+1)d + z_0)^2} 
\right)
\nonumber\\
&&- \frac{\pi \sigma_{S2}^2 (\alpha_{xx}^n (0) + \alpha_{yy}^n (0))}{2(\varepsilon_2 + 1)^2} 
\sum_{J=0}^\infty \sum_{K=0}^\infty
\Bigg(
\frac{\Delta_1^{J+K} \Delta_2^{J+K}}{(z_0 + (J+K)d)^2} 
- \frac{2 \, \Delta_1^{J+K+1} \Delta_2^{J+K}}{(J+K+1)^2 d^2} 
+ \frac{\Delta_1^{J+K+2} \Delta_2^{J+K}}{(z_0-(J+K+2)d)^2} 
\Bigg)
\nonumber\\
&&- \frac{\pi \sigma_{S2}^2 \alpha_{zz}^n (0)}{(\varepsilon_2 + 1)^2} 
\sum_{J=0}^\infty \sum_{K=0}^\infty
\Bigg(
\frac{\Delta_1^{J+K} \Delta_2^{J+K}}{(z_0 + (J+K)d)^2} 
+ \frac{2 \, \Delta_1^{J+K+1} \Delta_2^{J+K}}{((J+K+1)^2 d^2} 
+ \frac{\Delta_1^{J+K+2} \Delta_2^{J+K}}{(z_0 - (J+K+2)d)^2} 
\Bigg).   
\ea
Performing the double summation leads to Eq.~(\ref{EnSS}). 

\section{Proof that Eq.~(\ref{EnBB}) leads to Eq.~(\ref{dEsingleslab}) for $d\to\infty$}
\label{app:EnBB}
In this appendix, we show that Eq.~(\ref{EnBB}) leads to Eq.~(\ref{dEsingleslab}) for $d\to\infty$, holding $z_0$ fixed at a nonzero value. Starting from Eq.~(\ref{EnBB}), we have 
\ba
&&\delta E_n^{(BB)} 
\to 
- \frac{\pi \sigma_{B2}^2 (\alpha_{xx}^n (0) + \alpha_{yy}^n (0) + 2 \alpha_{zz}^n (0))}{2(\varepsilon_2 + 1)^2} 
\frac{1}{d} \Phi\left(\Delta_1\Delta_2, 1, \frac{z_0}{d}\right) 
\nonumber\\
&=& 
- \frac{\pi \sigma_{B2}^2 (\alpha_{xx}^n (0) + \alpha_{yy}^n (0) + 2 \alpha_{zz}^n (0))}{2(\varepsilon_2 + 1)^2} 
\sum_{J=0}^{\infty} \frac{(\Delta_1\Delta_2)^J}{z_0+Jd}
\nonumber\\
&=& 
- \frac{\pi \sigma_{B2}^2 (\alpha_{xx}^n (0) + \alpha_{yy}^n (0) + 2 \alpha_{zz}^n (0))}{2(\varepsilon_2 + 1)^2 z_0}
- \frac{\pi \sigma_{B2}^2 (\alpha_{xx}^n (0) + \alpha_{yy}^n (0) + 2 \alpha_{zz}^n (0))}{2(\varepsilon_2 + 1)^2} 
\sum_{J=1}^{\infty} \frac{(\Delta_1\Delta_2)^J}{z_0+Jd}.  
\nonumber
\ea
As we see, the second term in the last line vanishes as $d \to \infty$. 

\section{Proof that Eq.~(\ref{EnSS}) leads to Eq.~(\ref{shiftsingleslab}) as $d \to \infty$}
\label{app:EnSSproof}

As $z_0/d \to 0$ holding $z_0$ fixed at a nonzero value, the terms $\frac{1}{d^2} \Phi\left(\Delta_1\Delta_2,1,\frac{z_0}{d}\right) 
+ \frac{d-z_0}{d^3} \Phi\left(\Delta_1\Delta_2,2,\frac{z_0}{d}\right)$ in the fifth line of Eq.~(\ref{EnSS}) become divergent:  
\ba
&&\frac{1}{d^2} \Phi\left(\Delta_1\Delta_2,1,\frac{z_0}{d}\right) 
+ \frac{d-z_0}{d^3} \Phi\left(\Delta_1\Delta_2,2,\frac{z_0}{d}\right)
\nonumber\\
&=&
\frac{1}{d} \sum_{J=0}^{\infty} \frac{(\Delta_1\Delta_2)^J}{Jd + z_0}
+ \left( 1 - \frac{z_0}{d} \right)
\sum_{J=0}^{\infty} \frac{(\Delta_1\Delta_2)^J}{(Jd + z_0)^2}
\nonumber\\
&\underset{\text{$z_0/d \to 0$}}{\longrightarrow}&
\frac{1}{d} 
\frac{1}{z_0}
+ \left( 1 - \frac{z_0}{d} \right)
\frac{1}{z_0^2} = \frac{1}{z_0^2}.    
\ea
As the other terms are finite as $z_0/d \to 0$ for a fixed nonzero $z_0$, we have 
\be
\delta E_n^{(SS)} 
\underset{\text{$z_0/d \to 0$}}{\longrightarrow}
- \frac{\pi \sigma_{S2}^2 (\alpha_{xx}^n (0) + \alpha_{yy}^n (0) + 2 \alpha_{zz}^n (0))}{2(\varepsilon_2 + 1)^2 z_0^2}, 
\ee
and we recover our result Eq.~(\ref{shiftsingleslab}) for the charge disorder-induced energy level shift in an atom near a single slab. 

\end{widetext}


\begin{thebibliography}{99}
\bibitem{footnote-esu}
The conversion to SI units can be made using the fact that 1 C is $2.998\times 10^9$ esu. In SI units, the values of the bulk charge disorder variance per unit volume and the surface charge disorder variance per unit area would be $\sigma_B^2 = 1.3 \times 10^{-18} \, {\rm C^2/m^3}$ and $\sigma_S^2 = 3.5 \times 10^{-25} \, {\rm C^2/m^2}$.   
In cgs units, $e = 4.8032\times 10^{-10}\,{\rm esu}$, $\hbar = 1.0546\times 10^{-27} \, {\rm erg.s}$, and $c = 2.9979\times 10^{10} \, {\rm cm/s}$. We can check that these numbers give the right value for the fine-structure constant: $\alpha = e^2/(\hbar c) = 0.00730 \approx 1/137$.
\end{thebibliography}
\end{document}